\documentclass[10pt,journal,compsoc]{IEEEtran}

\ifCLASSOPTIONcompsoc
  \usepackage[nocompress]{cite}
\else
  \usepackage{cite}
\fi

\ifCLASSINFOpdf
\else
\fi
\usepackage[ruled,vlined]{algorithm2e}

\usepackage{comment}
\usepackage{multirow}
\usepackage{graphicx}
\usepackage{subfig}
\usepackage{amsmath}
\usepackage{hyperref}

\usepackage{textcomp}
\usepackage{colortbl}
\usepackage{hhline}

\usepackage{framed}
\usepackage{xcolor}

\newcommand{\tool}{\sc RMT}

\AtBeginDocument{%
  \providecommand\BibTeX{{%
    \normalfont B\kern-0.5em{\scshape i\kern-0.25em b}\kern-0.8em\TeX}}}
    
\begin{document}

\title{A Declarative Metamorphic Testing Framework for Autonomous Driving}



\author{Yao Deng, Xi Zheng, Tianyi Zhang, Huai Liu, Guannan Lou, Miryung Kim, Tsong Yueh Chen

\IEEEcompsocitemizethanks{\IEEEcompsocthanksitem Y. Deng, X. Zheng, G. Lou are with the Department
of Computing, Macquarie University, Sydney,
NSW.\protect
E-mail: yao.deng@hdr.mq.edu.au, james.zheng@mq.edu.au, lougnroy@gmail.com
\IEEEcompsocthanksitem T. Zhang is with the Department of Computer Science, Purdue University, West Lafayette, IN.\protect
E-mail: tianyi@purdue.edu
\IEEEcompsocthanksitem H. Liu, T.Y. Chen are with the School of Software and Electrical Engineering, Swinburne University of Technology, Melbourne, VIC.\protect
E-mail: [hliu, tychen]@swin.edu.au
\IEEEcompsocthanksitem M. Kim is with the Department of Computer Science, University of California, Los Angeles, Los Angeles, CA.
\protect
E-mail: miryung@cs.ucla.edu
}

\thanks{Manuscript received April 19, 2005; revised August 26, 2015.}}








\markboth{Journal of \LaTeX\ Class Files,~Vol.~14, No.~8, August~2015}%
{Shell \MakeLowercase{\textit{et al.}}: Bare Demo of IEEEtran.cls for Computer Society Journals}

\IEEEtitleabstractindextext{%
\begin{abstract}
Autonomous driving has gained much attention from both industry and academia. Currently, Deep Neural Networks (DNNs) are widely used for perception and control in autonomous driving. However, several fatal accidents caused by autonomous vehicles have raised serious safety concerns about autonomous driving models. Some recent studies have successfully used the metamorphic testing technique to detect thousands of potential issues in some popularly used autonomous driving models. However, prior study is limited to a small set of metamorphic relations, which do not reflect rich, real-world traffic scenarios and are also not customizable.  
This paper presents a novel declarative rule-based metamorphic testing framework called {\tool}. {\tool} provides a rule template with natural language syntax, allowing users to flexibly specify an enriched set of testing scenarios based on real-world traffic rules and domain knowledge. {\tool} automatically parses human-written rules to metamorphic relations using an NLP-based rule parser referring to an ontology list and generates test cases with a variety of image transformation engines. 
We evaluated {\tool} on three autonomous driving models. With an enriched set of metamorphic relations, {\tool} detected a significant number of abnormal model predictions that were not detected by prior work. Through a large-scale human study on Amazon Mechanical Turk, we further confirmed the authenticity of test cases generated by {\tool} and the validity of detected abnormal model predictions.
\end{abstract}
\begin{IEEEkeywords}
Metamorphic testing, Autonomous driving, testing
\end{IEEEkeywords}}

\maketitle
\IEEEdisplaynontitleabstractindextext

\IEEEpeerreviewmaketitle

\IEEEraisesectionheading{\section{Introduction}
\label{sec:introduction}}
\IEEEPARstart{A}{utonomous} driving has been through rapid development and has attracted increasing attention and investment from industry in recent years. 
In 2018, Waymo launched the first autonomous car service in the Phoenix metropolitan area, making one of the first steps towards commercializing autonomous vehicles~\cite{waymoS}.    
Deep Neural Networks (DNNs) are widely used to solve perception and control problems in autonomous driving~\cite{yang2018end, bojarski2016end}. 
However, recent traffic accidents caused by incorrect predictions of driving models have raised significant safety concerns. For instance, a crash caused by Tesla autopilot system led to the death of the driver, where 
the driving model failed to recognize the white truck against the bright sky~\cite{tesla}. 
Thus, it is crucial to detect erroneous behavior of driving models on various traffic scenarios to improve safety and robustness of autonomous driving.


A common practice to test autonomous vehicles in industry is through road test~\cite{zhang2016study, broggi2014proud}. However, road test is quite expensive. It is difficult to cover various weather conditions, road conditions, and driving scenes.
Simulation-based testing~\cite{zheng2019real, belanger2010and, dommel1969digital, sanchez1995variable, Dosovitskiy17, airsim2017fsr, yao2022scenario} is widely adopted to complement road test by mimicking driving scenarios in a simulated environment.
However, existing studies have questioned the fidelity of simulation and whether simulated driving scenarios can faithfully reflect real-world scenarios~\cite{stocco2021mind, zheng2015perceptions,lee2007computing,mitra2013verifying}.


Recent work has applied metamorphic testing (MT)~\cite{chen2018metamorphic, segura2016survey} to automatically synthesize new road images for testing driving models~\cite{zhang2018deeproad, tian2018deeptest}. These techniques rely on pre-defined metamorphic relations (MRs) between model predictions of an original image and a new image transformed from it. An MR example is, {\em the model prediction of steering angle should not change significantly after adding raindrops to a road image}. 

However, prior works~\cite{zhang2018deeproad, tian2018deeptest} are limited to hardcoded MRs for generating testing cases based on a small set of affine transformations. The limited MRs are not enough to guide a comprehensive generation of test cases and the evaluation of autonomous driving systems because they cannot cover complicated and diverse driving scenarios. To our best knowledge, there is no existing work that supports generating diverse MRs based on customized rules for constructing new test cases of autonomous driving systems. Recently, the generation of MRs is researched in other application domains. For example, Blasi et al.~\cite{blasi2021memo} proposed a tool called MeMo to automatically generate MRs based on Javadoc comments, which contains rich information, such as the summary of implemented functions. Rahman et al.~\cite{rahman2020mrpredt} also leveraged knowledge in Javadoc comments to create a dataset and then trained a text classification model to predict MRs. These work show that leveraging domain knowledge is a promising approach to reduce the effort of generating MRs. 

Along the same research direction, we propose to generate MRs for autonomous driving system testing based on domain knowledge from traffic rules or domain experts. For example, Table~\ref{tab:handbook} shows three traffic rules from driver handbooks.
These traffic rules describe the correct behavior of a driver when the driving environment changes, which could be converted to MRs.
In addition, domain experts can describe interesting scenarios to test or modify traffic rules according to actual situations in different regions. In this way, we can create diverse MRs based on different rules to generate test cases for different driving scenarios. Therefore, we design and develop a novel \textit{declarative \underline{R}ule-based \underline{M}etamorphic \underline{T}esting framework ({\tool})}. {\tool} allows testers to define and create testing rules in natural language. Testers can create their testing rules by referring traffic rules or other domain knowledge. Then {\tool} leverages a NLP-based semantic parser to extract grammar dependency predicates, identify elements to change in a driving scene by matching extracted predicates with a predefined ontology list, identify transformation to be applied using predicate translation, and create corresponding MR of the testing rule. {\tool} then generates new road images based on the ontology elements and transformations and then validates the correctness of model predictions based on the MR. Since driving scenarios based on testing rules require sophisticated image transformations such as adding or removing objects in an image, {\tool} makes use of several advanced computer vision techniques such as image semantic manipulation~\cite{wang2018high} and image-to-image translation~\cite{liu2017unsupervised} to support these transformations. {\tool} is also extensible to import new ontologies and image generation techniques to create more testing scenarios.

We evaluated {\tool} on three autonomous driving models that predict steering angle or driving speed. We experimented with seven testing rules to assess the usefulness of {\tool}. Given a dataset of 942 road images, {\tool} generated 3846 new images by applying the transformations induced by seven rules. Based on the metamorphic relations induced by the seven rules, {\tool} detected $2184$, $1314$, and $941$ abnormal predictions of these three driving models respectively. While these many abnormal predictions are detected, one may question whether they are indeed traffic rule violations. Prior work has never answered this question but only reported the number of detected abnormal predictions. To answer this question, for the first time, we conducted a large-scale human study with 64 drivers to assess the validity of detected abnormal model predictions on Amazon Mechanical Turk~\cite{mTurk}. The experiment shows the majority of detected abnormal model predictions are considered meaningful by human drivers. We also compared {\tool} with prior works~\cite{tian2018deeptest, zhang2018deeproad} and found that {\tool} is more effective to detect abnormal model predictions.

In summary, this work makes the following contributions:
\begin{itemize}
    \item  We, for the first time in its kind, proposed a declarative autonomous driving testing framework that allows users to flexibly specify metamorphic testing rules based on domain knowledge in natural language. An ontology was specifically defined to extract critical information from these rules. 
    
    \item We implemented the testing framework using ontology-based semantic parsing, image segmentation, and image translation techniques. Given the input testing rule written in natural language, {\tool} maximizes the automation for the testing and evaluation of autonomous driving systems, including the rule parsing, MR creation, and follow-up test case generation. The link of {\tool} prototype is \url{https://github.com/ITSEG-MQ/RMT-TSE}.
    
    
    
    \item We evaluated our framework on three autonomous driving models and demonstrated that our framework is capable of detecting a significant number of abnormal model predictions. We also conducted the first large-scale human study with 64 workers on Amazon Mechanical Turk to evaluate the validity of detected abnormal model predictions. 
    
\end{itemize}

\begin{table}[]
\centering
\caption{Traffic rule examples}
\label{tab:handbook}
\renewcommand{\arraystretch}{1.4}
\scalebox{0.98}{
\begin{tabular}{|l|l|}
\hline
\multicolumn{1}{|c|}{\textsf{Country/District}} & \multicolumn{1}{c|}{\textsf{Traffic rules}}                                                                                                                                                                                                \\ \hline
NSW, Australia~\footnote{\url{https://www.nsw.gov.au/sites/default/files/2021-05/road_users_handbook-english.pdf}}              & \begin{tabular}[c]{@{}l@{}}When you see potential hazards, slow down and \\ prepare to stop, for example when pedestrians are \\ close to the road or when other vehicles may turn \\ in front of you.\end{tabular}     \\ \hline
California, USA~\footnote{\url{https://www.dmv.ca.gov/portal/file/california-driver-handbook-pdf/}}      & \begin{tabular}[c]{@{}l@{}}A 3-sided red YIELD sign indicates that you must \\ slow down and be ready to stop, if  necessary, to \\ let any vehicle, bicyclist, or pedestrian pass  before \\ you proceed.\end{tabular} \\ \hline
Germany~\footnote{\url{https://adilbari.files.wordpress.com/2015/07/md-guide-to-driving-in-germany.pdf}}          & \begin{tabular}[c]{@{}l@{}}Drive more slowly at night because you cannot see\\ as far ahead and you will have less time to stop \\ for a hazard.\end{tabular}                                                       \\ \hline
\end{tabular}}
\vspace{-15pt}
\end{table}

\section{Related work}
\label{sec:related_work}

\subsection{Testing Autonomous Driving Systems}

\subsubsection{Image-based Testing}
Many works focused on generating test cases based on real-world driving images for testing autonomous driving models. DeepTest~\cite{tian2018deeptest} applied affine transformations such as rotation and blurring to generate test images. DeepRoad~\cite{zhang2018deeproad} applied a generative adversarial network (GAN) to create driving images in snowy or rainy weathers.  In~\cite{dreossi2017systematic}, Dreossi et al.~proposed to insert vehicles with different sizes into driving images to test CNN models for detecting and classifying vehicles. In this work, we propose to use three kinds of image generation techniques including image manipulation, image synnthesis, and image-to-image translation to construct new test images for different driving scenes that are derived from traffic rules.

On the other hand, some works applied adversarial attacks~\cite{deng2021deep} to generate test images that look similar to original driving images but can cause driving models make wrong predictions. In~\cite{pei2017deepxplore}, DeepXplore was proposed to generate driving images that maximize the neuron coverage of the driving model under test using a optimization-based adversarial attack method. In~\cite{wicker2018feature}, Wicker et al.~proposed an adversarial attack method to add perturbations on the most vulnerable pixels in a image. Changes of such pixels would affect predictions of a CNN model for traffic sign detection. In~\cite{zhou2018deepbillboard}, Deepbillboard was proposed to replace normal billboards with adversarial ones in driving images to test the robustness of driving models. Though adversarial attack-based methods can easily generate new driving images to expose misbehaviors of driving models, such images may not be realistic because adversarial attacks need to either directly modify images collected from cameras or modify objects such as traffic signs on road. In this work, we mainly focus on the generation of driving images that can occur in real world and may cause faults of driving models.

\subsubsection{Simulation-based Testing}
Besides generating driving images, several recent works create driving scenarios in simulation environments to test autonomous driving systems. In~\cite{gambi2019generating}, Gambi et al.~proposed to use NLP techniques to extract information from police reports and then reconstruct driving scenarios based on extracted information. In this study, we propose an ontology to describe traffic scenes, use NLP techniques to extract ontology elements from human defined rules, create metamorphic relations, and generate test driving images. Several works~\cite{abdessalem2018testing, ben2016testing, gambi2019automatically, riccio2020model} proposed to use search algorithms to generate critical driving scenarios that cause collision or deviation of the autonomous vehicle in the simulation environment. In their work, they adopted domain knowledge to design objective functions such as time to collision (TTC) to help find critical driving scenarios. In our work, we leverage domain knowledge to design MRs and new driving scenes. The MRs can also be applied in simulation environment to create driving scenarios. The main difference is to use a simulator to render driving scenarios instead of applying image generation techniques on real-world driving images. We leave this as a future work. In~\cite{riccio2020model}, Riccio and Tonella applied human study to assess whether generated images are recognizable from human's view. In our work, we applied human study to evaluate the authenticity of generated images and the validity of detected abnormal model predictions.

\subsection{Testing and Debugging of Deep Learning Models}
Recently, many techniques have been proposed for testing and debugging deep learning models~\cite{pei2017deepxplore, ma2018deepgauge, sun2018concolic, ma2018deepmutation,  guo2018dlfuzz, du2019deepstellar,  kim2019guiding, xie2019deephunter,  lee2020effective, ma2018mode, ren2020few, tao2020trader, zhang2019apricot}. For example, Pei et al.~\cite{pei2017deepxplore} developed a white-box testing framework called DeepXplore, which optimizes neuron coverage to generate test inputs for activating previously uncovered neurons in a DL model. DeepGauge~\cite{ma2018deepgauge} extended neuron coverage and proposed multi-granularity coverage for DL models.  Different from the white-box optimization-based methods, our proposed method is a black-box method to generate meaningful testing images based on traffic rules, without the knowledge such as neuron values of driving models. 
 DeepConcolic~\cite{sun2018concolic} leverages concolic testing to generate adversarial examples for DL models. Lee et al.~\cite{lee2020effective} proposed an adaptive neuron-selection strategy to select most vulnerable neurons in DL models to improve the testing coverage and efficiency. MODE~\cite{ma2018mode} facilitates DL debugging by leveraging differential analysis to find faulty neurons in CNNs. TRADER~\cite{tao2020trader} leverages trace divergence analysis and embedding regulation to debug RNNs.

There is also a large body of verification techniques for deep learning models~\cite{pulina2010abstraction, katz2017reluplex, huang2017safety, wang2018efficient, singh2018fast, paulsen2020neurodiff, paulsen2020reludiff, li2020prodeep, du2020marble}. 
Pulina et al.~\cite{pulina2010abstraction} proposed an abstraction-refinement approach to examine the safety of a neural network. 
Reluplex~\cite{katz2017reluplex} leveraged SMT solving to verify the robustness of DNNs with the ReLU activation function. 
Huang et al.~\cite{huang2017safety} proposed a framework to find adversarial examples using SMT and exhaustive search.

\subsection{Metamorphic Testing} 
Chen et al.~first proposed metamorphic testing (MT) to address the oracle problem in software testing~ \cite{chen1998metamorphic}. Given source test cases with corresponding outputs, MT transforms original test inputs to generate new test inputs and validates the program output on a new input based on metamorphic relations (MRs). MRs specify the  relationships between the outputs of the original input and a new input. When test results violate MRs, the program has a bug. For example, suppose a program \(f\) implements the SINE function. We can define an MR, \(f(x) = f(x + 2\pi)\). Then we can generate a new test case \(x + 2\pi\) and test whether the output of the original test input x (i.e., \(f(x)\)) and the output of \(x + 2\pi\) (i.e., \(f(x +2\pi)\)) are equal~\cite{chen2016metric}. MRs are also able to reveal relationships between two outputs beyond equality. An example is for an accommodation booking system.
The searching result of rooms with filtering conditions (e.g., the price of a room is in a certain range) should be a subset of that without filtering conditions. In this study, we proposed an MT testing framework that allows users to define their MRs in natural language to test autonomous driving systems.

MT has been widely applied to many domains such as  middleware~\cite{chan2006integration}, healthcare~\cite{bojarczuk2021measurement}, and machine learning~\cite{xie2011testing}. For example, Murphy et al.~\cite{murphy2008properties} proposed six MRs including additive, multiplicative, permutative, invertive, inclusive, and exclusive to transform input data for support vector machines. Based on this work, Xie et al.~\cite{xie2011testing} further proposed MRs to test specific supervised ML models such as K-Nearest Neighbor (KNN) classifiers and Naive Bayes classifiers. In~\cite{zhou2019metamorphic}, Zhou et al.~applied MT to test Lidar-based perception system in an autonomous driving system called Apollo. DeepTest~\cite{tian2018deeptest} and DeepRoad~\cite{zhang2018deeproad} are representative approaches that apply MT to autonomous driving models. However, these two approaches only support equality MRs. In other words, the outputs of an original input and a new input should be the same or similar within a threshold. In addition, prior works (e.g., DeepTest~\cite{tian2018deeptest} and DeepRoad~\cite{zhang2018deeproad}) were limited to often hardcoded rules. In this work, we propose a testing framework that allows users to specify testing rules, e.g., a car should slow down if a stop sign is added to the curbside. Then the framework automatically parses input testing rules using NLP techniques and uses the corresponding ontology to generate MRs beyond equality.  Furthermore, our framework is integrated with more image transformations than DeepTest and DeepRoad to support other MRs. 

\subsection{Generation of MRs}
The generation of MRs is the core task for MT. However, it is difficult and challenging to propose a general or universal MR generation method to automatically generate MRs. Prior work~\cite{tian2018deeptest, zhang2018deeproad, zhou2019metamorphic} proposed hardcoded MRs to guide the generation of testing cases and model evaluation. Recently, some works proposed MR generation methods for some application domains by using search-based techniques or using ML techniques to predict MRs. A tool called GAssertMRs~\cite{ayerdi2021generating} was proposed to automatically generate MRs for cyber-physical systems based on genetic programming. Blasi et al.\cite{blasi2021memo} proposed MeMo to automatically generate equivalent MRs from Javadoc comments using NLP techniques. In addition, Rahman et al.~\cite{rahman2020mrpredt} trained a text classification model to predict MRs based on Javadoc comments. In this study, we proposed the generation of MRs by parsing input testing rules originating from domain knowledge using NLP techniques, a proposed ontology, and predicate inference. The declarative approach of MR generation based on testing rules in natural language can support diverse and complicated testing scenarios in autonomous driving.

\section{The RMT Framework}
\label{sec:methodology}

\begin{figure*}[ht]
\centering
\includegraphics[width = 1\textwidth]{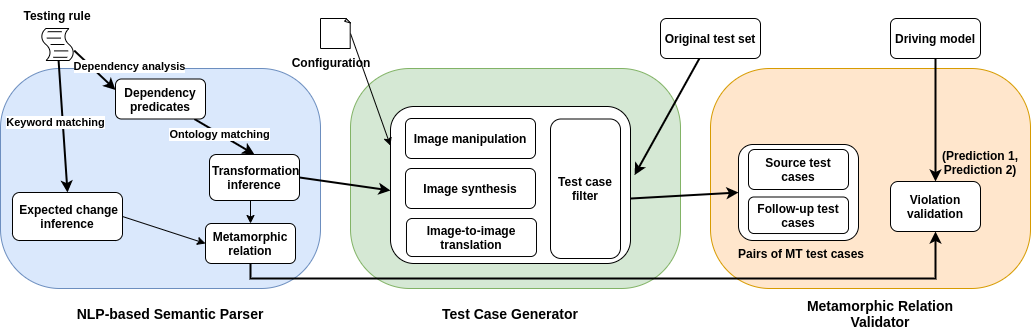}
\vspace{-20pt}
\caption{An overview of {\tool}}
\vspace{-6pt}
\label{fig:framework-workflow}
\end{figure*}

{\tool} allows users to interactively and flexibly define custom metamorphic driving scenarios for testing autonomous driving models. Figure \ref{fig:framework-workflow} shows the architecture of {\tool}. It consists of three components: (1) an NLP-based semantic parser that induces metamorphic relations (MRs) from human-written test scenarios, (2) a test generator that generates new road images based on the induced MR, and (3) a validator that detects MR violations. The testing process is semi-automated. When the human-written test rules are provided, the tool can automate the process of test case generation  driving model evaluation.

First, a user describes a testing rule following the IFTTT (If-This-Then-That) paradigm~\cite{mi2017empirical} in natural language (Section~\ref{sec:specification}). The testing rule specifies how to change a driving scenario as well as the expected change of the driving behavior. Then an NLP-based semantic parser leverages a driving scenario ontology to match and extract key information in the testing rule (Section~\ref{sec:parser}). Then, an MR is established based on the extracted information. Based on this MR and a configuration file, the test generator invokes the corresponding image generation technique to generate new road images (Section~\ref{sec:MT generator}). 
For each pair of the original and the newly generated images, the violation validation is processed to check whether the model predictions satisfy the MR (Section~\ref{sec:detect}). If an MR is violated, an abnormal prediction is detected, indicating a violation of the human-written testing rule. Section~\ref{sec:configration} demonstrates the implemented prototype and describes how to flexibly configure and modify the driving scenario ontology and the test generator. The user of {\tool} needs to provide a testing rule, configurations of image transformation engines, and original test cases (i.e., road images) as input.


\subsection{Testing Rule Specification}
\label{sec:specification}

{\tool} accepts testing rules described in natural language based on the IFTTT  paradigm. The IFTTT rule syntax contains one or more {\em if-then} statements. In the {\em if} clause, a user describes how to change a driving scenario. In the {\em then} clause, the user describes the expected change in driving behaviors caused by the scenario change. 
For example, we can specify a testing rule {\em  ``If a pedestrian appears on the roadside, then the ego-vehicle should slow down at least 30\%''}. The {\em if} clause specifies that a driving scenario can be changed by adding a pedestrian on the roadside. The {\em then} clause specifies the expected behavior change, which is the speed of the ego-vehicle should decrease at least 30\%.  We chose to use IFTTT paradigm because it is suitable to cover most driving scenarios in traffic rules that describes correct human behaviors when specific driving conditions are met. IFTTT paradigm may not be able to describe some complicated driving scenarios such as collisions involving multiple vehicles. However, such driving scenarios are not described in traffic rules and thus beyond the scope of this paper.

\subsection{NLP-based Semantic Rule Parsing}
\label{sec:parser}

{\tool} applies an NLP-based semantic parser to extract information in the IFTTT rule and create the corresponding MR. From the {\em if} clause, the semantic parser identifies  information including the element to change in the driving scene, the transformation to be applied on the element, and how to apply the transformation (i.e., transformation parameters). From the {\em then} clause, the parser extracts the expected change of the driving behavior.

Section~\ref{sec:dependency} describes how {\tool} applies Part-of-Speech (POS) tagging and grammar dependency analysis to analyze the structure of a testing rule and build a grammar dependency graph. Section~\ref{sec:ontology} describes how to identify the element to change by matching elements in the grammar dependency graph with a predefined ontology. Section~\ref{sec:infer1} describes how to infer the metamorphic transformation and transformation parameters based on extracted key elements in the {\em if} clause. Section~\ref{sec:infer2} describes how to infer the expected change based on the extracted key elements in the {\em then} clause. Section~\ref{sec:nested} describes how to handle dynamic scenarios with a sequence of rules.

\subsubsection{Dependency Analysis}
\label{sec:dependency}

\begin{figure}[ht]
\centering
\includegraphics[width = 0.4\textwidth]{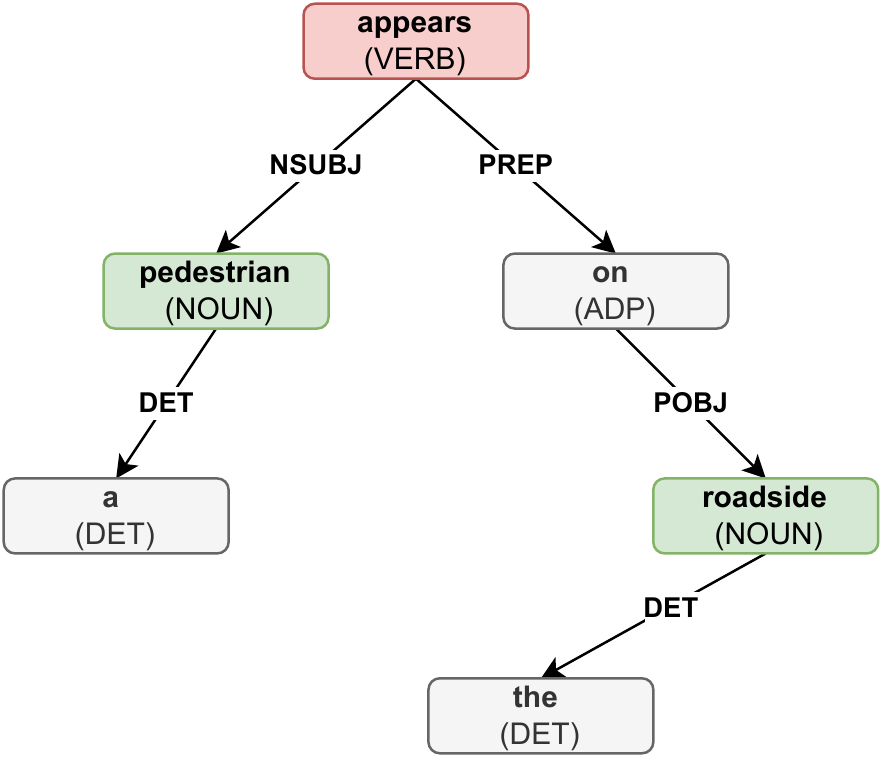}

\caption{An example of dependency parsing}

\label{fig:dependency}
\end{figure}

Given a testing rule, the parser first splits the rule into {\em if} and {\em then} clauses. Then, for each clause, the parser applies part-of-speech (POS) tagging~\cite{schmid1994part} to identify the part of speech of each word based its definition and context. The parser then uses grammar dependency analysis~\cite{yamada2003statistical} to identify the grammatical relationships between words and generate a dependency graph.   
For example, given a sentence, ``{\em a pedestrian appears on the roadside}'', Figure~\ref{fig:dependency} shows the resulting dependency graph and POS tags such as VERB and NOUN. The dependency {\em nsubj} means {\em pedestrian} is the {\em nominal subject} of {\em appears}. Table~\ref{tab:depdendency} shows common grammar dependencies in a sentence.

\begin{table}[]
\centering
\caption{Common grammar dependencies in a sentence}
\label{tab:depdendency}
\renewcommand{\arraystretch}{1.2}
\begin{tabular}{|l|l|}
\hline
\multicolumn{1}{|c|}{\textsf{Grammar Dependency}} & \multicolumn{1}{|c|}{\textsf{Description}}              \\ \hline
NSUBJ                       & Nominal subject                   \\ \hline
DOBJ                        & Direct object                     \\ \hline
NMOD                        & Nominal modifier                  \\ \hline
AMOD                        & Adjectival modifier               \\ \hline
ADVMOD                      & Adverb modifier                   \\ \hline
NPADMOD                     & Noun phrase as adverbial modifier \\ \hline
DET                         & Determiner                        \\ \hline
\end{tabular}
\end{table}

\begin{table*}[]
\centering
\caption{The Ontology of Traffic Scenes}
\label{tab:ontology}
\renewcommand{\arraystretch}{1.2}
\scalebox{1}{
\begin{tabular}{|l|l|l|l|}
\hline
 \multicolumn{1}{|c|}{\textsf{Category}}                    & \multicolumn{1}{|c|}{\textsf{Level-1 Subcategory}}            & \multicolumn{1}{|c|}{\textsf{Level-2 Subcategory}} & \multicolumn{1}{|c|}{\textsf{Properties}}                                                                                                                               \\ \hline
\multirow{7}{*}{Road network}        & \multirow{4}{*}{Road part}              & Lane                         & \begin{tabular}[c]{@{}l@{}}direction: {[}forward, reverse{]}\\ orientation: {[}vertical, horizontal{]}\\ position: {[}left, right{]}\end{tabular} \\ \cline{3-4} 
                                     &                                         & Line                         & \begin{tabular}[c]{@{}l@{}}Type: {[}solid, dash{]}\\ Color: {[}white, yellow{]}\end{tabular}                                                      \\ \cline{3-4} 
                                     &                                         & Crosswalk                    & Orientation: {[}vertical, horizontal{]}                                                                                                           \\ \cline{3-4} 
                                     &                                         & Sidewalk                     &  ---                                                                                                                                                 \\ \cline{2-4} 
                                     & \multirow{3}{*}{Traffic infrastructure} & Traffic sign                 & \begin{tabular}[c]{@{}l@{}}Type: {[}stop, speed limit, turn{]}\\ Shape: {[}circle, square{]}\end{tabular}                                         \\ \cline{3-4} 
                                     &                                         & Traffic light                & Color: {[}red, yellow, green{]}                                                                                                                   \\ \cline{3-4} 
                                     &                                         & Stop and yield line                    & ---                                                                                                                                                  \\ \hline
\multirow{5}{*}{Object}              & \multirow{2}{*}{Static object}          & Tree                         & ---                                                                                                                                       \\ \cline{3-4} 
                                     &                                         & Building                     & ---                                                                                                                                      \\ \cline{2-4} 
                                     & \multirow{3}{*}{Dynamic object}         & Pedestrian                   & \begin{tabular}[c]{@{}l@{}} ---\\ \end{tabular}                                               \\ \cline{3-4} 
                                     &                                         & Vehicle                      & \begin{tabular}[c]{@{}l@{}}Type: {[}car, truck, van, school bus{]}\\ Color: {[}white, black, blue, ...{]}\end{tabular}                   \\ \cline{3-4} 
                                     &                                         & Bicyclist                    &  ---                                                                                                                                  \\ \hline
\multirow{2}{*}{Environment} & {Weather}                & rainy, cloudy, snowy                        &   Level: [light, normal, heavy]                                                                                                                                                                                 
                                                                                                            \\ \cline{2-4} 
                                     & {Time}                   & day, night                          & ---                                                                                                                                                                                                                                                                                       \\ \hline
\end{tabular}
}
\end{table*}

\subsubsection{Ontology-based Information Extraction}
\label{sec:ontology}

Ontology~\cite{studer1998knowledge} is a methodology to formally summarize categories and create a knowledge base to organize properties and relations in a specific domain. In this work, we define a driving scenario ontology to model elements in driving scenarios. Specifically, we define an ontology containing three categories of elements, as shown in Table~\ref{tab:ontology}. The first category contains elements for forming a road network such as lanes, lines, and crosswalk, and traffic infrastructure such as traffic signs, and traffic lights. The second category includes objects in a driving scenario such as pedestrians and vehicles. The third category is about weather and driving time. The ontology is inspired by the five-layer driving scene ontology proposed in~\cite{bagschik2018ontology}. We combine the first three layers in~\cite{bagschik2018ontology} to the category Road network and keep the rest two layers as two categories Object and Environment. We combine the first three layers  in~\cite{bagschik2018ontology} because they are all related to road network elements including the topology and geometry of roads, traffic infrastructure on roads, and temporary manipulation of road elements. Then, we enrich each category with elements and properties. In~\cite{bagschik2018ontology}, Bagschik et al.~used their ontology to model hundreds of driving scenarios for German motorways, which has justified the ability of the ontology to describe diverse driving scenarios. An ontology element can contain specific properties. For example, a vehicle has type and color properties. For some ontology elements, we predefined some properties and their values as shown in Table~\ref{tab:ontology}. For these categories Road network, Object and Environment, ontology elements in the level-2 subcategory will be matched in the input testing rule. 

Given the POS tags from the previous step, the semantic parser identifies all nouns and matches them with elements in the ontology using WordNet Wu-Palmer (WUP) similarity~\cite{pedersen2004wordnet}. WordNet is a lexical database of semantic relations including  synonyms, hyponyms, and meronyms between words. WUP similarity is used to measure the semantic similarity between a pair of words based on WordNet database. The similarity score is between $(0, 1]$ where $1$ means two words have identical semantic meaning. Specifically, we use the WUP similarity API in a Python library Natural Language Toolkit (NLTK)~\cite{bird2009natural} to calculate the similarity between each pair of extracted noun and ontology element. If the similarity score is above a threshold ($0.75$ in this paper), the noun is considered matched with the ontology element. The threshold is set as 0.75 based on the experiment result in a previous study~\cite{liu2006application} where the threshold of 0.75 helped achieve the optimal F1 score. In the previous example, the noun ``pedestrian'' is identified and matched with the ontology element {\em pedestrian} with the similarity of $1$. If the extracted noun from the input rule is ``person'', it will also be matched with the ontology element {\em pedestrian} because the similarity between two words is $0.89$, which is higher than the threshold.

When an ontology element is matched, the parser further identifies its properties. 
 The parser first checks the grammar dependency graph and extracts dependencies starting from the ontology element (i.e., the noun is the head node in the dependency graph) as predicates, where the dependency name is the predicate name. For example, for the description {\em a black car}, {\em DET(a, car)}, {\em AMOD(black, car)} can be extracted but only the dependency {\em AMOD(black, car)} is kept because it shows ``black'' is an adjective to modify the ontology element ``car''. Finally, the parser matches extracted adjectives from predicates with predefined property values using  WordNet similarity. In the example, ``black'' is matched with the property ``{\em color}'' of the ontology element ``car''.

\subsubsection{Inferring Metamorphic Transformation}
\label{sec:infer1}
In this work, we define three transformations to change an ontology element in a driving scenario, including {\em Add, Remove,} and {\em Replace}. A new ontology element such as a car can be {\em added} to a traffic scene. An existing element such as a traffic sign can be {\em removed} from a driving scenario. Furthermore, an element can be replaced by another element. For example, a white car can be replaced with a black car, and the sunny day can be replaced with the rainy day. For different transformations, they have different parameters. The remove transformation only takes a target ontology element as an input parameter. The {\em add} transformation takes a target ontology element, a reference ontology element, and a position predicate, such as {\em on} and {\em front}, to describe the relative location with respect to the reference element. The {\em replace} transformation requires a target ontology element and a new element as parameters.

To identify the transformation described in the {\em if} clause of a testing rule, the parser first extracts the root verb from the grammar dependency graph and matches it with the three predefined transformations using word2vec~\cite{mikolov2013efficient}. Word2vec is a method to learn word associations and convert words to vector representations. Words that are used in similar context or have similar meaning are close to each other in the vector representation space. Therefore, we can use the distance between two word vectors to measure the similarity between two words. In this work, we use this method to match extracted verb and pre-defined transformations. First, we calculate the distance between the verb and all pre-defined transformations using a pre-trained word2vec model in a Python library SpaCy~\cite{spacy}, which can be downloaded in the link~\footnote{\url{https://spacy.io/models/en}}. Then, we match the verb to the transformation within the shortest distance in the vector representation space. We use word2vec but not WordNet for transformation match because word2vec performs better to match verbs that implicitly present same meanings but are not synonyms in our pilot study. 

In the previous example, the verb ``appears'' is matched with {\em Add}. Thus, a proposition, \texttt{TRANSFORMATION(``appears'', {\em add})}, is generated accordingly. Then, the subject of this verb is extracted to produce another proposition in the form of \texttt {NSUBJ(noun, verb)}, such as \texttt {NSUBJ(``pedestrian'', ``appears'')} in the previous example. The object of the verb is also extracted accordingly in the form of \texttt {DOBJ(noun, verb)}. In some cases, a verb does not have a direct object but a prepositional phrase. In such cases, the parser extracts the transit dependency by traversing the dependency graph to find the noun in the prepositional phrase and uses the preposition as the predicate. In the previous example, \texttt{ON(``roadside'', ``appears'')} is extracted from the prepositional phrase ``on the roadside.'' 

Table~\ref{tab:predicate} describes the logic rules to infer a transformation from the generated propositions. Take the first infer rule of {\em Add} as an example. \texttt {NSUBJ($n_1$, v)} means the predicate \texttt {NSUBJ} should have been extracted from the testing rule. \texttt {ONTOLOGY($n_1$)} means the noun extracted from the proposition \texttt {NSUBJ($n_1$, v)} should be matched as an ontology element. \texttt {TRANSFORMATION(v, add)} means the extracted verb should be matched as the add transformation. \texttt{ON($n_2$, v)} means the predicate \texttt {ON} should have been extracted and \texttt{ONTOLOGY($n_2$)} means the noun in the proposition \texttt{ON($n_2$, v)} should be matched as an ontology element. When all of these conditions are met, \texttt{ADD($n_1$, $n_2$, on)} is inferred, which means add $n_1$ on $n_2$.  More examples of testing rules that can be parsed by transformation inference rules can be found in Table~\ref{tab: rules} (e.g., ''change`` is inferred to the transformation {\em Replace} in Rule 7).

\begin{table*}[]
\centering
\caption{Transformation Inference Rules}
\label{tab:predicate}
\renewcommand{\arraystretch}{1.4}

\begin{tabular}{|l|l|}

\hline
\multirow{3}{*}{\textbf{Add}}     & NSUBJ(\(n_1\), v) \(\wedge\) ONTOLOGY(\(n_1\)) \(\wedge\) TRANSFORMATION(v, add) \(\wedge\) ON(\(n_2\), v) \(\wedge\) ONTOLOGY(\(n_2\)) \(\rightarrow\) ADD(\(n_1\), \(n_2\), on)                                                                                                                      \\ \cline{2-2} 
                                  & NSUBJ(\(n_1\), v) \(\wedge\) ONTOLOGY(\(n_1\)) \(\wedge\) TRANSFORMATION(v, add) \(\wedge\) FRONT(\(n_2\), v) \(\wedge\) ONTOLOGY(\(n_2\)) \(\rightarrow\) ADD(\(n_1\), \(n_2\), front)                                                                                                                \\ \cline{2-2} 
                                  & NSUBJ(\(n_1\), v) \(\wedge\) ONTOLOGY(\(n_1\)) \(\wedge\) TRANSFORMATION(v, add) \(\wedge\) BEHIND(\(n_2\), v) \(\wedge\) ONTOLOGY(\(n_2\)) \(\rightarrow\) ADD(\(n_1\), \(n_2\), behind)                                                                                                              \\ \hline
\textbf{Remove}                   & NSUBJ(n, v) \(\wedge\) ONTOLOGY(n) \(\wedge\) TRANSFORMATION(v, remove) \(\rightarrow\) REMOVE(n)                                                                                                                                                                                    \\ \hline
\multirow{2}{*}{\textbf{Replace}} & \begin{tabular}[c]{@{}l@{}}NSUBJ(\(n_1\), v) \(\wedge\) ONTOLOGY(\(n_1\)) \(\wedge\) TRANSFORMATION(v, replace) \(\wedge\) PREP(\(n_2\), v) \(\wedge\) ONTOLOGY(\(n_2\)) \(\wedge\) \\ WEATHER (\(n_2\)) \(\wedge\) WEATHER(\(n_2\)) \(\rightarrow\) REPLACE(\(n_1\), \(n_2\), weather)\end{tabular} \\ \cline{2-2} 
                                  & \begin{tabular}[c]{@{}l@{}}NSUBJ(\(n_1\), v) \(\wedge\) ONTOLOGY(\(n_1\)) \(\wedge\) TRANSFORMATION(v, replace) \(\wedge\) PREP(\(n_2\), v) \(\wedge\) ONTOLOGY(\(n_2\)) \(\wedge\) \\ OBJECT (\(n_1\)) \(\wedge\) OBJECT(\(n_2\)) \(\rightarrow\) REPLACE(\(n_1\), \(n_2\), object)\end{tabular}    \\ \hline
\end{tabular}
\end{table*}

\subsubsection{Inferring Expected Change}
\label{sec:infer2}

\begin{table}[]
\centering
\caption{Expected Change Inference Rules}
\label{tab:mr}
\renewcommand{\arraystretch}{1.4}
\scalebox{0.8}{
\begin{tabular}{|l|}
\hline
CHANGE(decrease) \(\rightarrow\) \(x_1>x_2\) \\ \hline

CHANGE(increase) \(\rightarrow\) \(x_1<x_2\) \\ \hline

CHANGE(decrease) \(\wedge\) NEGATION(description) \(\rightarrow\) \(x_1<=x_2\) \\ \hline

CHANGE(increase) \(\wedge\) NEGATION(description) \(\rightarrow\) \(x_1>=x_2\) \\ \hline

\begin{tabular}[c]{@{}l@{}} CHANGE(decrease)  \(\wedge\) (MODIFIER(at least) \(\lor\) MODIFIER(more than))\\ \(\wedge\) QUANTITY (n, number) \(\rightarrow\) \(x_1-x_2>=n\)\end{tabular} \\ \hline

\begin{tabular}[c]{@{}l@{}} CHANGE(decrease)  \(\wedge\) (MODIFIER(at least) \(\lor\) MODIFIER(more than))\\ \(\wedge\) QUANTITY (n, number) \(\wedge\) NEGATION(description) \(\rightarrow\) \(x_1-x_2<=n\)\end{tabular} \\ \hline

\begin{tabular}[c]{@{}l@{}} CHANGE(increase)  \(\wedge\) (MODIFIER(at least) \(\lor\) MODIFIER(more than))\\ \(\wedge\) QUANTITY (n, number) \(\rightarrow\) \(x_2-x_1>=n\)\end{tabular} \\ \hline

\begin{tabular}[c]{@{}l@{}} CHANGE(increase)  \(\wedge\) (MODIFIER(at least) \(\lor\) MODIFIER(more than))\\ \(\wedge\) QUANTITY (n, number) \(\wedge\) NEGATION(description) \(\rightarrow\) \(x_2-x_1<=n\)\end{tabular} \\ \hline

\begin{tabular}[c]{@{}l@{}} CHANGE(decrease) \(\wedge\) (MODIFIER(at least) \(\lor\) MODIFIER(more than))\\ \(\wedge\) QUANTITY (n, percentage) \(\rightarrow\) \((x_1-x_2)/x_1>=n\%\)\end{tabular} \\ \hline

\begin{tabular}[c]{@{}l@{}} CHANGE(decrease) \(\wedge\) (MODIFIER(at least) \(\lor\) MODIFIER(more than))\\ \(\wedge\) QUANTITY (n, percentage) \(\wedge\) NEGATION(description) \(\rightarrow\) \((x_1-x_2)/x_1<=n\%\)\end{tabular} \\ \hline

\begin{tabular}[c]{@{}l@{}} CHANGE(decrease) \(\wedge\) (MODIFIER(less than) \\   \(\wedge\) QUANTITY (n, number) \(\rightarrow\) \(x_1-x_2<=n \wedge x_1>x_2\)\end{tabular} \\ \hline

\begin{tabular}[c]{@{}l@{}} CHANGE(decrease) \(\wedge\) (MODIFIER(less than) \\   \(\wedge\) QUANTITY (n, number) \(\wedge\) NEGATION(description) \(\rightarrow\) \(x_1-x_2>=n\)\end{tabular} \\ \hline

\begin{tabular}[c]{@{}l@{}} CHANGE(decrease) \(\wedge\) MODIFIER(less than) \\   \(\wedge\) QUANTITY (n, percentage) \(\rightarrow\) \((x_1-x_2)/x_1<=n\% \wedge x_1>x_2\)\end{tabular} \\ \hline

\begin{tabular}[c]{@{}l@{}} CHANGE(decrease) \(\wedge\) MODIFIER(less than) \\   \(\wedge\) QUANTITY (n, percentage) \(\wedge\) NEGATION(description) \(\rightarrow\) \((x_1-x_2)/x_1>=n\%\)\end{tabular} \\ \hline

\begin{tabular}[c]{@{}l@{}} CHANGE(increase) \(\wedge\) (MODIFIER(less than) \\   \(\wedge\) QUANTITY (n, number) \(\rightarrow\) \((x_2-x_1)<=n \wedge x_2>x_1\)\end{tabular} \\ \hline

\begin{tabular}[c]{@{}l@{}} CHANGE(increase) \(\wedge\) (MODIFIER(less than) \\   \(\wedge\) QUANTITY (n, number) \(\wedge\) NEGATION(description) \(\rightarrow\) \(x_2-x_1>=n\)\end{tabular} \\ \hline

\begin{tabular}[c]{@{}l@{}} CHANGE(increase) \(\wedge\) MODIFIER(less than) \\   \(\wedge\) QUANTITY (n, percentage) \(\rightarrow\) \((x_2-x_1)/x_1<=n\% \wedge x_2>x_1\)\end{tabular} \\ \hline

\begin{tabular}[c]{@{}l@{}} CHANGE(increase) \(\wedge\) MODIFIER(less than) \\   \(\wedge\) QUANTITY (n, percentage) \(\wedge\) NEGATION(description) \(\rightarrow\) \((x_2-x_1)/x_1>=n\%\)\end{tabular} \\ \hline

CHANGE(same) \(\rightarrow\) \(|x_1>x_2| <= \delta \) \\ \hline
\end{tabular}}
\end{table}

In this work, we consider three possible {\em expected changes}, including {\em increase} ({\em left} for steering angles), {\em decrease} ({\em right} for steering angles), and {\em staying the same}. Furthermore, a change can be described by a {\em change modifier}, such as {\em at least} and {\em more than}, or a {\em change quantity}, such as a number and a percentage. 

As the description of the {\em then} clause is often simple, the parser directly matches key information without performing dependency analysis. The parser first applies POS tagging and Name Entity Recognition (NER) on the {\em then} clause. Then, it matches the expected change from verbs or nouns, the change modifier (if any) from adjectives or adverbs, and the change quantity (if any) from numerals. Specifically, we define a lexicon for expected changes for each type of behavior, such as ``increase'', ``decrease'', ``same'', ``accelerate'', ``slow'', ``deviate'', etc. Then the parser matches extracted verbs with expected change lexicons using WUP similarity. In the previous example of ``{\em the vehicle should slow down}'', ``{\em slow}'' is matched with the {\em decrease} behavior. Therefore, the parser infers the expected change as {\em decrease} and generates a proposition \texttt{CHANGE(decrease)}. 


To identify the {\em change modifier}, the parser first identifies adjectives and adverbs in the {\em then} clause. Then the parser checks the adjective as well as its neighbor words together and matches the phrase with a  predefined lexicon of change extent including {\em at least}, {\em more than} and {\em less than}. In the example of {\em then the ego-vehicle should slow down at least 30\%}, `least'' is first identified and then ``at least'' is found and matched with the change extent {\em at least}. Finally, a proposition, \texttt{MODIFIER(at least)}, is generated.

The parser also considers the condition of negation words such as ``no'', ``not''. When these words occur in a sentence, the meaning of described change modifier should be reversed, such as from ``more than'' to ``no more than''. Therefore, the parser matches the occurrence of negation words. When a negation word is matched, a proposition \texttt{NEGATION(description)} is generated.    

The change quantity is extracted from numeral words. The parser applies Name Entity Recognition (NER)~\cite{mohit2014named} to check whether there is a number or a percentage in the {\em then} clause. In the example of {\em then the ego-vehicle should slow down at least 30\%}, ``30\%'' is found and identified as a percentage. A proposition, \texttt{QUANTITY(``30'', percentage)}, is then generated.

When all propositions are generated, the parser uses them to infer an expected change function \(E\).
Table~\ref{tab:mr} describes the logic rules to infer an expected change function from the generated propositions,  where $x_1$ and $x_2$ refer to model predictions on the original image and the generated image. Take the first rule as an example. If no change modifier or change quantity propositions are generated, the expected change formula is simply defined as \(x_1>x_2\) or \(x_1<x_2\). 
If a change modifier proposition and a change quantity proposition are generated, the formula is defined with respect to these two values. For example, if there are two propositions, \texttt{MODIFIER(``least'', at least)} and \texttt{QUANTITY(``30'', percentage)}, then a formula,  $(x_1-x_2)/x_1>=30\%$, is generated. If the {\em expected change} is {\em same}, the formula is defined as \(|x_1-x_2| <= \delta \), meaning the difference between two model predictions should be smaller than a threshold $\delta$.  If a negation word occurs, the comparison operator will be reversed accordingly.

After the transformation function and the expected change formula are obtained, {\tool} uses them to create an MR. Let the original driving image be \(x_o\), the transformation function be \(T\) and its parameters be \(p\), the test generator be \(G\), the driving model under test be \(F\), and the expected change formula be \(E\). The expected change \(E\) could be $0$ or within a small range (e.g., $[-0.01, 0.01]$) if the MR represents an equality relation. Then the new driving image is represented as \(G(x_o, T, p)\), and model predictions on two driving images are \(F(x_o)\) and \(F(G(x_o, T, p))\). The generated MR is shown as Formula~\ref{eq:mr}, which describes that the change of the input driving image causes the change of the model prediction. 
\begin{equation}
\label{eq:mr}
  G(x_o, T, p) \rightarrow E(F(x_o), F(G(x_o, T, p)))
\end{equation}

\subsubsection{Testing Dynamic Scenarios with a Sequence of Rules}
\label{sec:nested}
Our framework also supports combing two or more IFTTT blocks in one rule to test dynamic scenarios. For example, an autonomous vehicle should understand the risk of hitting a pedestrian with respect to its distance to the pedestrian. Therefore, a tester may want to check whether the ego-vehicle slows down more when the pedestrian is closer to it. In the case, we can write a testing rule, ``{\em If a pedestrian appears on the roadside, the speed should decrease. If he gets closer to the vehicle, the speed should decrease more.}'' This rule contains two IFTTT blocks. The first block ``{\em If a pedestrian appears on the roadside, the speed should decrease.}'' describes the basic test scenario and the expected change is the comparison of model predictions on the original image \(x_o\) and generated image \(g_1\). The second block describes a subsequent scenario where another image \(g_2\) is generated based on the original image. The added pedestrian in $g_2$ should be closer to the ego-vehicle. The expected change between model predictions on \(g_1\) and \(g_2\) is also defined in the second block.

To support such rules indicating the comparison between model predictions of two driving scenarios, the parser first applies dependency analysis and POS tagging on both two IFTTT blocks to extract key information as describe in the previous sections. From the first IFTTT block, the transformation function and expected change function are created. For the second block, if the parser cannot find the same ontology elements described in the first block, the parser uses a pronoun resolution technique~\cite{hobbs1977pronoun} to match pronouns with ontology elements. In the example ``{\em If a pedestrian appears on the roadside, the speed should decrease. If he gets closer to the vehicle, the speed should decrease more.}'', ``he'' is extracted and matched with the ontology element {\em pedestrian} in the first {\em if} clause.

Furthermore, the parser checks comparative adjectives or adverbs occurred in the second IFTTT block and matches them with comparative adjectives and adverbs such as {\em closer}, {\em more}, {\em faster}, and {\em slower}. The comparative adjective or adverb in the second {\em if} clause is the additional parameter in the transformation proposition. In the example {\em ``If he gets closer to the vehicle''}, {\em closer} is identified as a comparative adjective and it is added to the transformation proposition obtained from the first block. The new transformation propostion of the second block is thus \texttt{Add(pedestrian, sidewalk, on, closer)}. The comparative adjective or adverb in the second {\em then} clause indicates the comparison of model predictions. In the example {\em ``the speed should decrease more''}, {\em decrease} and {\em more} are extracted, which infers to the expected change inequity \(x_2>x_3\) where \(x_2\) is the model prediction on the generated image from the first IFTTT block and \(x_3\) is generated from the second IFTTT block.

For the rule containing two IFTTT blocks, two MRs are created and both of them should be satisfied, as shown in Formula~\ref{eq:mr3}. It expresses the condition to compare model predictions on two generated images. 

\begin{equation}
\label{eq:mr3}
  \left\{
  \begin{aligned}
      g_1 = G(x_o, T, p_1) \rightarrow E(F(x_o), F(g_1)) \\
      g_2 = G(x_o, T, p_2) \rightarrow E(F(g_1), F(g_2))
  \end{aligned}
  \right.
\end{equation}

\subsection{Metamorphic Test Generator}
\label{sec:MT generator}

Given the inferred metamorphic transformation identified in the previous step, the test generator creates new driving images using three different image generation techniques---image manipulation, image synthesis, and image-to-image translation. To make the image generation process more extensible, we parameterize the test generator and allow users to supplement new image generation techniques through a configuration file (detailed in Section~\ref{sec:configration}). Essentially, this configuration file specifies the command to invoke a transformation technique and a mapping between the transformation technique and transformation propositions identified in the previous step.

\subsubsection{Image Manipulation}
\begin{figure*}
\centering
\subfloat[The original image]{\includegraphics[width=.3\textwidth]{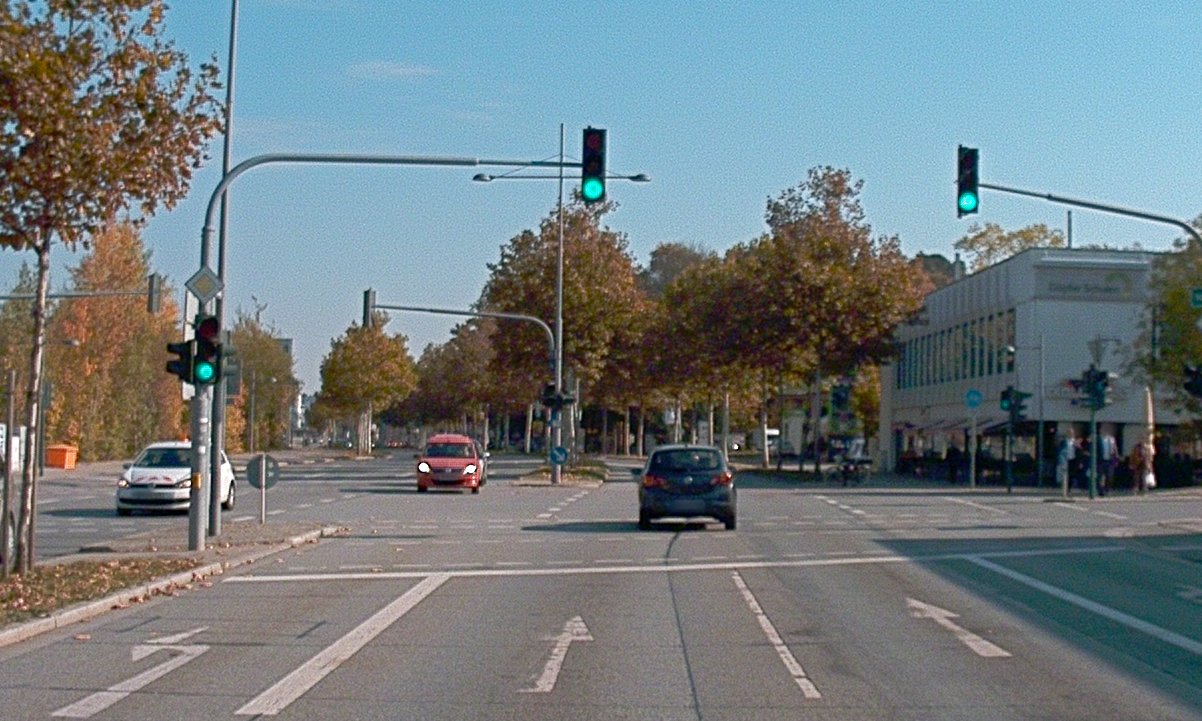}\label{fig:rule5_1}}\hfil
\subfloat[The image after segmentation]{\includegraphics[width=.3\textwidth]{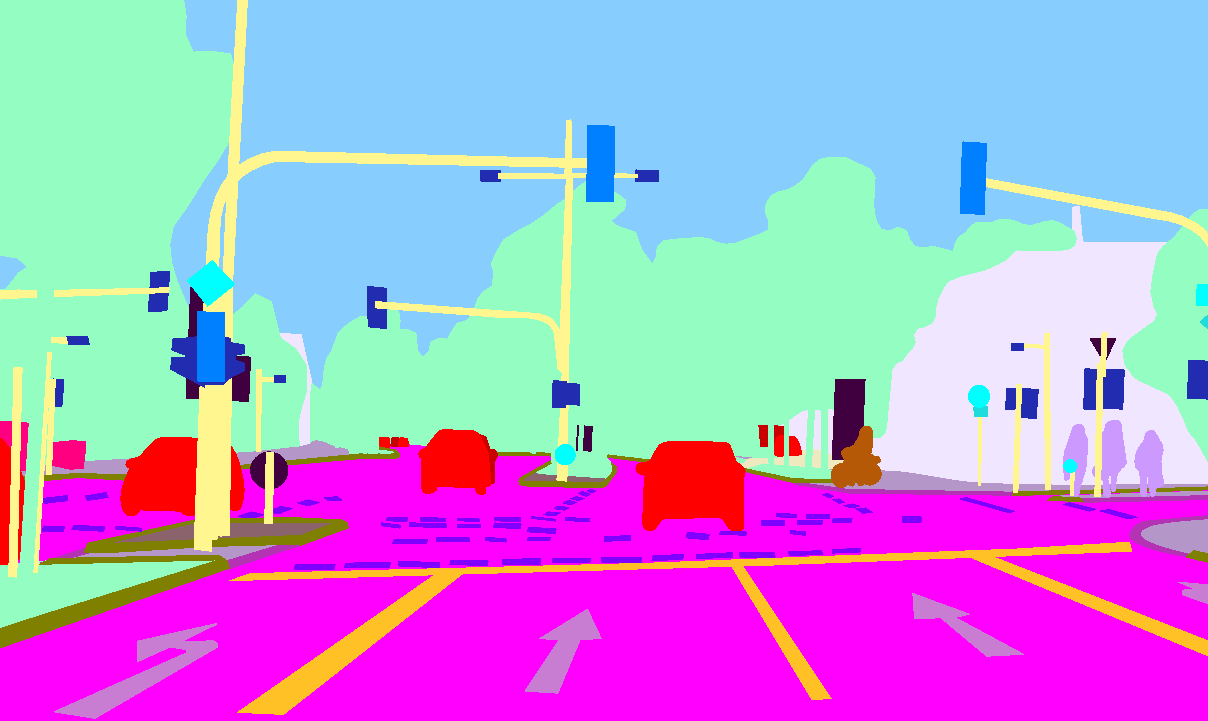}\label{fig:rule5_2}}\hfil
\subfloat[The segmented vehicle]{\includegraphics[width=.3\textwidth]{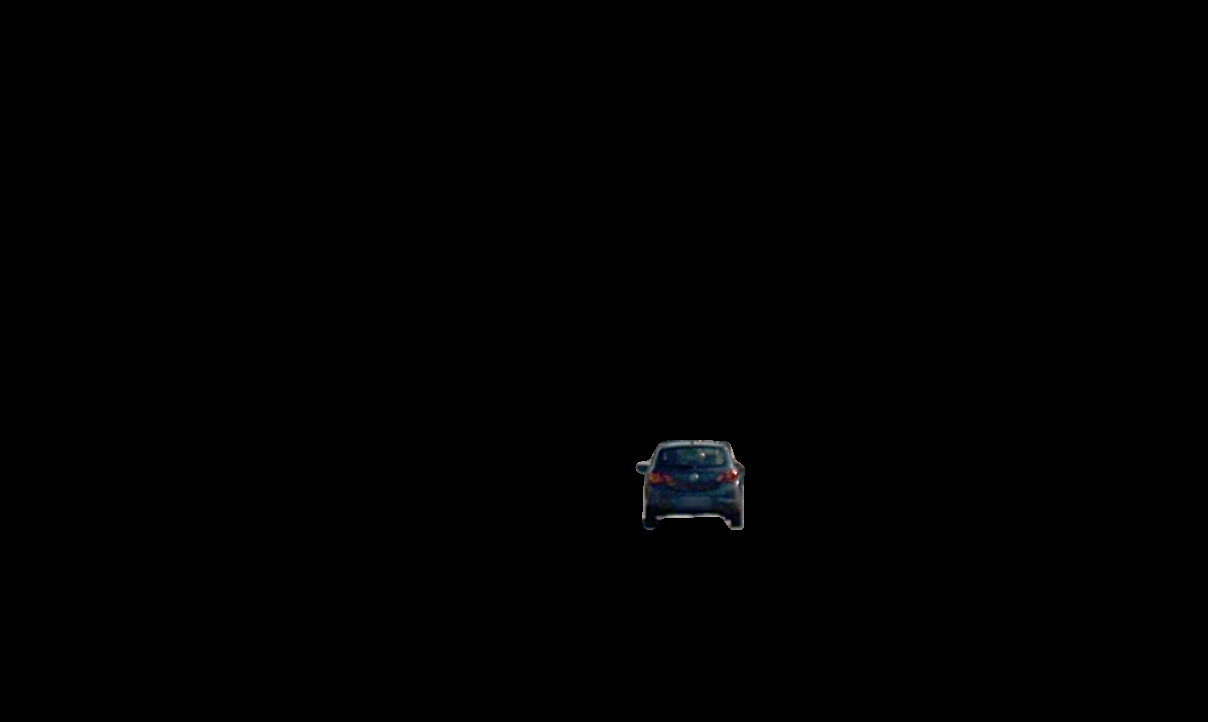}\label{fig:rule5_2}}\hfil 
\caption{An example of image segmentation}\label{figure}
\vspace{-6pt}
\label{fig:semantic_label_map}
\end{figure*}

We develop an image manipulation technique to directly add ontology elements in a road image. This is done through semantic label maps. A semantic label map provides image classification results in pixel level. As Figure~\ref{fig:semantic_label_map} shows, image (a) is the driving scene, and image (b) is the semantic label map associating each pixel in the image (a) with a class of different color (e.g., the blue area in the semantic label map represents the sky in the original image).  Based on the semantic label map, we are able to extract an object (i.e., also known as a {\em mask} in computer vision) belonging to a class from the original image, as shown in image (c). We use OpenCV~\cite{opencv_library} to identify the exact position of different objects and extract object masks as templates. In this work, we use a driving dataset with  annotated semantic label maps. Semantic label maps can be automatically generated using semantic segmentation models such as DeepLab~\cite{deeplabv3plus2018}. We leave the integration of such techniques as future work. 

By navigating through the semantic label maps in the original driving dataset, we create a gallery of object images for each ontology element. To add an ontology element into driving images, {\tool} first matches the ontology element with its template mask in the gallery. Then {\tool} adds the mask into the driving images using OpenCV. In the previous example, the ontology element is {\em a pedestrian} and the transformation parameter is {\em roadside}. Therefore, the corresponding template mask is selected and added into test images. In this step, one challenge is that for different road images,  the position of roadside is very likely to be different. We solve this challenge by checking the boundary of a road using the semantic label map. More specifically, we move the mask horizontally by changing the x-axis coordinates of the mask and then check whether at the current position of the mask (i.e., coordinates in the driving image where the mask is moved to), the semantic label of the left bottom pixel is changing from road to other class' semantic label. If so, we add the mask at that position. 

Mathematically, let a driving image with height \(h\) and width \(w\); let the bottom left point of the image be the coordinate origin; let an ontology element mask with height \(h'\) and width \(w'\), the coordinate of the mask's bottom left point be \((x, y)\), \(y+h'<=h\) and \(x+w'<=w\); and let the semantic map of the image is \(M\) and \(M(p_x, p_y)\) is the semantic class label of the point \(p_x, p_y\). We identify the position of the mask as \(x+x', y\), which satisfy conditions:
\begin{equation}
\label{eq:add_mask}
  \left\{
  \begin{aligned}
      M(x+x'-1, y) = road \\
      M(x+x', y) \neq road \\
      x+x'+w'<=w
  \end{aligned}
  \right.
\end{equation}

To add a pedestrian closer to the vehicle, we just need to firstly move the mask from \((x, y)\) to \((x, y-y')\) and use above rules to identify the add position.

\subsubsection{Image Synthesis}
\begin{figure*}
\centering
\subfloat[Original driving scene]{\includegraphics[width=.16\textwidth]{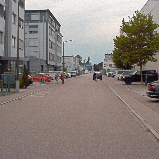}\label{fig:generate_a}}\hfil
\subfloat[Original driving scene]{\includegraphics[width=.16\textwidth]{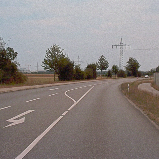}\label{fig:generate_b}}\hfil 
\subfloat[Original driving scene]{\includegraphics[width=.16\textwidth]{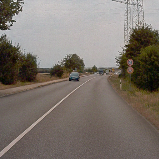}\label{fig:generate_d}}\hfil 
\subfloat[Original driving scene]{\includegraphics[width=.16\textwidth]{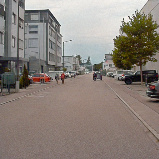}\label{fig:generate_e}}\hfil 
\subfloat[Original driving scene]{\includegraphics[width=.16\textwidth]{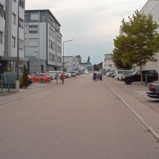}\label{fig:generate_c}}\hfil 

\subfloat[Add a pedestrian] {\includegraphics[width=.16\textwidth]{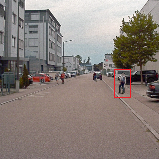}\label{fig:generate_f}}\hfil   
\subfloat[Add a slow sign]{\includegraphics[width=.16\textwidth]{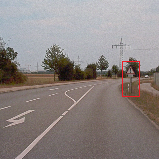}\label{fig:generate_g}}\hfil
\subfloat[Remove a lane line]{\includegraphics[width=.16\textwidth]{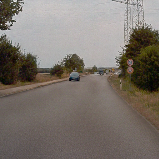}\label{fig:generate_i}}\hfil
\subfloat[Replace buildings with trees]{\includegraphics[width=.16\textwidth]{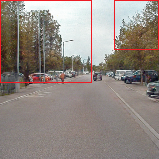}\label{fig:generate_j}}\hfil
\subfloat[Transform to a night scene]{\includegraphics[width=.16\textwidth]{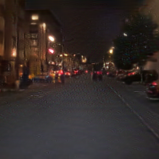}\label{fig:generate_h}}\hfil
\caption{Examples of source images and transformed images for Rules 1,2,3,5,6,7}\label{figure}
\vspace{-6pt}
\label{fig:transformed_images}
\end{figure*}

Though image manipulation technique can implement {\em Add} transformation fast with low cost, it falls short in several cases. First, the added element sometimes may look unnatural since it is extracted from another image with a different driving environment. Second, it cannot support {\em Remove} and {\em Replace} transformations. Therefore, we develop an image synthesis technique based on a generative model called {\em Pix2pixHD}~\cite{wang2018high}. {\tool} applies the model to remove and replace ontology elements. Pix2pixHD is a Generative Adversarial Network (GAN)~\cite{goodfellow2014generative} for image generation based on semantic label maps. Given a semantic label map of a driving scene image, {\em Pix2pixHD} synthesizes the corresponding driving scene image. For example, if the ontology element is ``dashed lane lines'' and the transformation is ``Remove'', the test generator detects pixels in the semantic label map with the class as dashed lines and then change the label of these pixels to road. The transformation effect is shown as from Figure~\ref{fig:generate_c} to Figure~\ref{fig:generate_h} Similarly, if the user wants to replace the buildings in the driving images with trees, the test generator changes the pixels of buildings in the semantic label to trees. Then the test generator invokes {\em Pix2pixHD} to generate a new driving image based on the modified semantic label map (i.e., from Figure~\ref{fig:generate_d} to Figure~\ref{fig:generate_i}).

\subsubsection{Image-to-Image Translation}
Even though the generative model-based image synthesis can replace objects by manipulating semantic label maps, it cannot be applied to ontology elements without specific shapes such as weather and time. Therefore, we adopt another technique image-to-image translation to solve this problem. We use an image-to-image translation GAN called {\em UNIT}~\cite{liu2017unsupervised}. The main function of {\em UNIT} is to transform images from one domain to another (e.g., changing a tiger to a lion). We integrate pre-trained models of {\em UNIT} on  BDD100K~\cite{yu2018bdd100k},  which is an autonomous driving dataset containing a large amount of driving images in different weather conditions and driving time.

\subsubsection{Image Filtering}
When generating new driving images, not all testing images are qualified to be transformed. For example, if the user wants to add a vehicle in the front of the ego-vehicle but there is already a vehicle or other objects at the same position, such images should be filtered out. Therefore, we develop a test case filtering mechanism in {\tool} to automatically filter invalid images. For object addition, {\tool} checks whether another object also exists at the position to put the mask. For object removal or replacement, {\tool} checks whether the size of the object is too small. If an object is very small, removing or replacing the object is unlikely to make an influence on the model prediction. Though recent research found that adversarial examples that have slightly modification on original images would make driving models produce wrong prediction, this paper aims to generate meaningful new test images (test generation) not for robustness testing.
New images generated by removing or replacing small objects are rather relevant to the robustness testing, and thus not considered in this study.  
For weather and time replacement, {\tool} filters generated driving images that are similar to original driving images within the bounds of the mean squared error (MSE), since in this condition the original driving images are most possibly already in the target weather or driving time.     

\subsection{Metamorphic Relation Validator}
\label{sec:detect}
Given a metamorphic testing set with source test cases (i.e., original driving images) and follow-up test cases (i.e., generated new driving images), the metamorphic relation validator feeds each test case into the driving model under test and obtains model predictions. Then, the validator uses the created MR to validate each prediction pair (prediction $x_1$ for a source test and prediction $x_2$ for the corresponding follow-up test). If the prediction pair violates the MR, an abnormal model prediction is detected. If the testing rule contains two IFTTT blocks, a prediction tuple ($x_1$, $x_2$, $x_3$) is fed to the validator and two created MRs are evaluated. Both two MRs should be satisfied simultaneously, otherwise a violation is said to be detected. 



\subsection{Implementation}

\label{sec:configration}
\begin{figure}[ht]
\centering
\includegraphics[width = .47\textwidth]{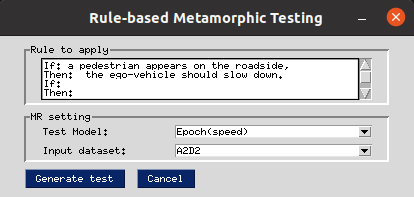}
\caption{The main interface for specifying traffic rules in {\tool}}
\label{fig:gui_main}
\end{figure}

\begin{figure}[ht]
\centering
\includegraphics[width = .45\textwidth]{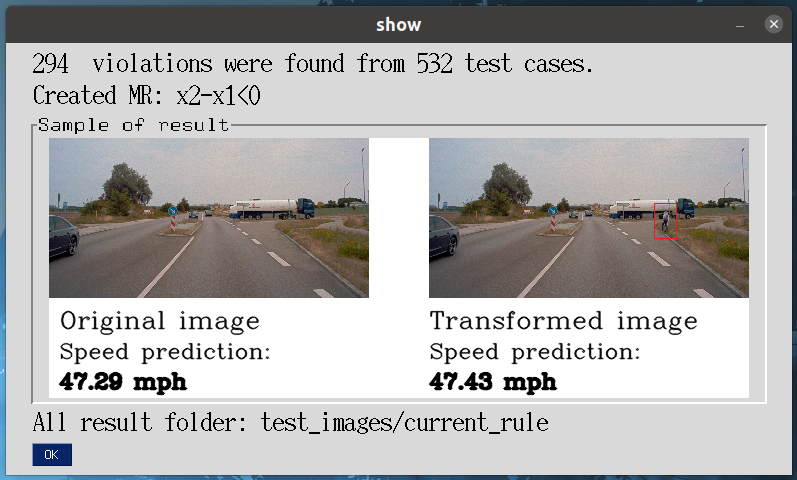}
\caption{A sample of detected traffic rule violations for user inspection. A pedestrian is added on the roadside while the ego-vehicle does not slow down.}
\label{fig:gui_result}
\end{figure}

We implemented a prototype of {\tool} in Python and PyTorch.
Figure~\ref{fig:gui_main} shows the main GUI of the prototype. A model tester can specify the driving rule to apply in the text field and then select the driving model and the original test set. The prototype can automatically parse the rule, create the MR, generate source and follow-up test cases, and detect violations. A separate window (Figure~\ref{fig:gui_result}) will pop up to show the metamorphic testing result. It first shows the number of detected violations and the total number of generated test cases (e.g., $294$ violations were found out of $532$ test cases). A sample of detected violations is rendered in the middle of the popup window. In this example, a pedestrian is added in the new image, while the driving speed does not decrease, thus the violation. By comparing the predictions, the tester can verify whether the MR is indeed violated. Finally, a line of text at the bottom shows the storage path of images for all detected violations.

\begin{figure}[ht]
\centering
\includegraphics[width = .5\textwidth]{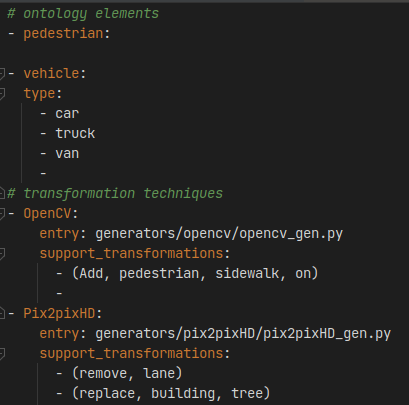}
\caption{A configuration example}
\label{fig:config}
\end{figure}

{\tool} uses a YAML configuration file to manage and coordinate components in the {\tool} framework. Specifically, the configuration file stores pre-definend ontology elements and their properties, the transformation list, and the invocation command of each image generation technique. Figure~\ref{fig:config} shows a subsection of a configuration file. In the example, we define ontology elements including lane, building, and tree, transformations including remove and replace. We then configure a image generation model Pix2pixHD to implement transformations. For the image generation model, the attribution {\em entry} describes how to call the model and the attribution {\em support\_transformations} defines what transformation with its specific ontology parameters can be implemented by the generation model. Such setting allows flexible extension of {\tool}. For instance, we can enrich the ontology list by adding new elements and their parameters into the configuration file. If we implement a new image generation model, we can add it into the configuration file as well and specify its supportable transformations and parameters.   

\section{Experimental Settings}
\label{sec:experiment}

We conducted two experiments to evaluate the effectiveness of {\em RMT} and answered the following research questions:

\begin{itemize}
    
    \item [RQ1] How many abnormal model predictions can {\tool} detect with the seven proposed rules?

    \item [RQ2] How authentic are new road images generated by {\tool}?
    
    \item [RQ3] How valid are the abnormal model predictions detected by {\tool}?

    \item [RQ4] How does {\tool} perform compared with existing techniques (e.g. DeepTest and DeepRoad)?
    
    \item [RQ5] Is {\tool} efficient to generate testing images and test driving models?

\end{itemize}



\begin{table}
\centering
\caption{Proposed rules}
\label{tab: rules}
\scalebox{1.1}{
\begin{tabular}{l} 
\hline
\begin{tabular}[c]{@{}l@{}}\textbf{Rule 1}\\If: a pedestrian appears on the roadside,\\Then: the ego-vehicle should slow down.~\end{tabular}  \\
\begin{tabular}[c]{@{}l@{}}\textbf{Rule 2} \\
If: a speed limit sign appears on the roadside,\\
Then: the ego-vehicle should slow down.\end{tabular}  \\
\begin{tabular}[c]{@{}l@{}}\textbf {Rule 3} \\
If: a pedestrian appears on the roadside,\\
Then: the ego-vehicle should slow down at least 30\%.~\end{tabular}  \\
\begin{tabular}[c]{@{}l@{}}\textbf{Rule 4} \\
If: a pedestrian appears on the roadside,\\
Then: the ego-vehicle should slow down. \\
If: he gets closer to the ego-vehicle,\\
Then: the speed should decrease more.~\end{tabular}  \\
\begin{tabular}[c]{@{}l@{}}\textbf{Rule 5} \\
If: lane lines are removed from the road, \\
Then: the steering angle of ego-vehicle should keep the same.~\end{tabular}  \\
\begin{tabular}[c]{@{}l@{}}\textbf{Rule 6} \\
If: the buildings are replaced with trees, \\
Then: the steering angle of ego-vehicle should keep the same.~\end{tabular}  \\
\begin{tabular}[c]{@{}l@{}}\textbf{Rule 7} \\
If: the driving time changes into night,\\
Then: the ego-vehicle should slow down.~\end{tabular}  \\
\hline
\end{tabular}}
\end{table}









For RQ1, we proposed seven rules as shown in Table~\ref{tab: rules} to test three driving models. We then measured how many violations were detected by MRs generated from these rules. To design such rules, we read traffic rule handbooks and pick up rules that contain supportable transformations and ontology elements by {\tool} prototype. Rules 1, 2, and 7 are directly derived from traffic rules from official driver handbooks in three different countries by rephrasing the rules in IFTTT syntax, as shown in Table~\ref{tab:handbook}. These rules check whether a driving model can handle potential hazards (e.g., a pedestrian may cross the road), recognize traffic signs, and recognize different driving scenes (e.g., daytime vs.~nighttime). Rules 3-6 are custom rules. Rule 3 is a customized from Rule 1 with a specific deceleration threshold 30\%, which can be set flexibily based on user expectation. We further evaluated all rules with different thresholds from 0 to 50\% for speed and steering angle. Rule 4 extends the traffic rule of NSW in Table~\ref{tab:handbook} to express a more complicated scenario---if the ego-vehicle is getting closer to a pedestrian, it should decelerate more.  Rules 5 and 6 are not derived from traffic rules, but they demonstrate how users can specify their own interesting driving scenes by adding, removing, or replacing objects in a road image. We further designed more complicated rules based on Rules 1-7 to compare the performances of simple rules and complicated rules.

We used A2D2 dataset~\cite{geyer2020a2d2} to train autonomous driving models for speed and steering angle predictions. 
The dataset contains $41,227$ images collected from 15 different locations and times. These images are stored in 15 folders. We used images collected in 13 folders as the training set to train autonomous driving models for steering and speed predictions, one folder as the validation set, and the last one as the test set containing 942 images.  We used A2D2 in our experiment because this dataset provides rich labels including semantic label maps, speed, and steering angles needed in transformation engines and training autonomous driving models. When training models for steering angle prediction, we converted the source label ``steering wheel angle" in the dataset to ``steering angle" based on the steering ratio 14.4:1, because the steering wheel angle from 0 to 360 degrees corresponds to the turn of vehicle wheels (i.e., ``steering angle'') from 0 to 25 degrees. As A2D2 dataset does not provide the parameter of steering ratio, we followed the same setting of the dataset released by Udacity~\cite{udacity_steer}.  Finally, the range of steering angles is $[-25, 25]$, where negative degrees mean that the vehicle is turning left. 

For Rules 1, 2, 4, and 7, the threshold $\epsilon$ was set as 0 by default. For Rule 3, the threshold $\epsilon$ was set as 30\% as described in the rule. For Rules 5 and 6, the threshold $\epsilon$ was set as 1.39. In~\cite{zhang2018deeproad}, the tolerance setting started from 10 degrees. However, in the previous study~\cite{zhang2018deeproad}, the range of steering wheel angle is [-180, 180] while in our paper the range of steering angle is [-25, 25]. We thus scaled the value of $\epsilon$ accordingly to a small value to reflect a similar standard (1.39$\simeq$10/(180/25)). Note that our experiment results do not show much differences when $\epsilon$ becomes larger.

We implemented and trained three CNN-based autonomous driving models for the evaluation: Epoch~\cite{epoch}, Resnet101~\cite{he2016deep}, , and VGG16 \cite{simonyan2014very}. Epoch is a top-performing self-driving model released by Udacity. The driving model architecture is adapted from Nvidia Dave-2~\cite{bojarski2016end} and achieves better performance in Udacity Challenge~\cite{Udacity}. We chose Epoch to represent driving models including Chauffeur~\cite{chauffeur} proposed in Udacity challenge because Epoch has the simplest architecture while achieves similar performances. ResNet is a state-of-the-art CNN architecture with residual blocks as components. VGG16 is a widely used architecture for general-purpose computer vision tasks. We adapted both ResNet and VGG16 into regression driving models by replacing the classifiers in two models as a  three-layer feed-forward network where the last layer contains $1$ neuron to predict speeds or steering angles.

For these driving models, we standardized their input image size to $320\times 160$. We chose the input size as $320\times 160$ which has the same scale as the input dataset and does not distort the resized driving scene. For the images from {\em A2D2}, we first cropped the center part from the original size of 
$1920\times 1208$ and then resized the images to $320\times 160$. Specifically, we removed the top $248$ rows of an image to make the size as $1920 \time 960$ and then resized it to $320\times 160$. We kept most part of the original image to ensure that the added objects such as pedestrians on the roadside would not be removed if we only keep the center crop of an image. For all autonomous driving models, we used Adam Optimizer with the default learning rate $0.0001$. The error rates of these driving models on the {\em A2D2} were measured by Mean Absolute Error (MAE), which is a common metric to measure the performance of regression models. The MAEs of three driving models for steering angle prediction with Epoch, VGG16, and Resnet101 are $2.68$, $2.65$, and $2.73$ respectively. The MAEs of three driving models for speed prediction are $3.09$, $3.30$, and $3.02$, respectively. 

We also used the test set to train the transformation engine {\em Pix2pixHD}, which ensures that {\em Pix2pixHD} can generate authentic follow-up test sets that look similar to the original test set. Since we used the default network architecture and hyper-parameter setting of {\em Pix2pixHD}, it will generate transformed images with size $1024\times 512$.  To fit the input requirement for three driving models under test, the generated images were cropped and resized to $320\times 160$. 
As introduced in Section~\ref{sec:MT generator}, each transformation engine has a built-in filter to select images that are applicable for transformation in the original test set. For seven proposed rules, transformation engines generated seven sets of metamorphic group of test cases with $532$, $425$, $532$, $335$, $476$, $604$, and $942$  pairs (source test cases, follow-up test cases) respectively.  



\begin{figure}[ht]
\centering
\includegraphics[width = .47\textwidth]{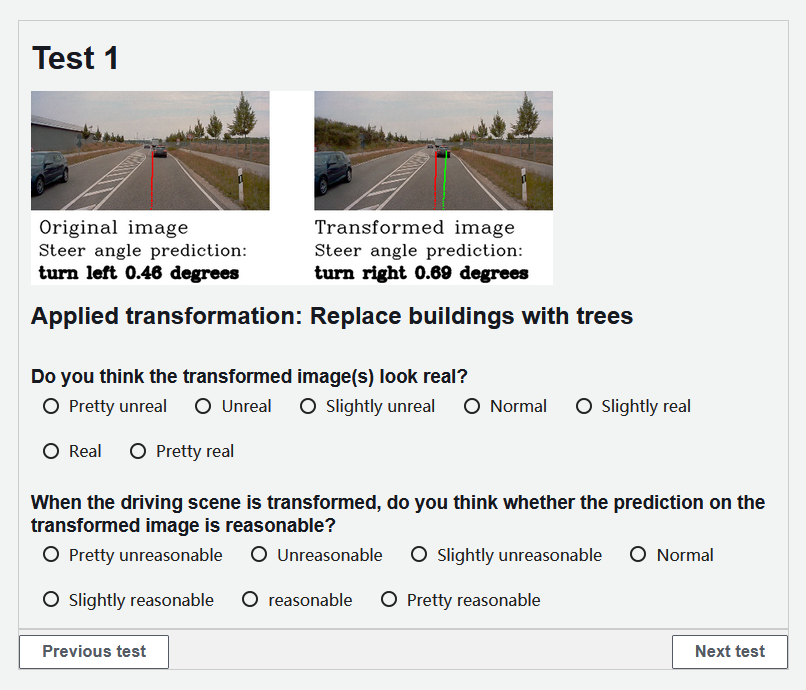}
\caption{An example test in a task on Amazon Mechanical Turk}
\label{fig:mturk}
\end{figure}

For RQ2 and RQ3, we published an evaluation task on a crowdsourcing platform, Amazon Mechanical Turk (mTurk)~\cite{mTurk}. An example test in a evaluation task is shown in Figure~\ref{fig:mturk}. We sampled test image pairs---a source test image and 
a corresponding transformed test image---as well as the corresponding model predictions of VGG16. We selected VGG16 as the target model in this experiment, since {\tool} detected abnormal model predictions using all seven rules.  We asked the workers to view the test 
image pairs and rate (1) whether the 
transformed image looks real and (2) 
whether the model prediction on the transformed image is reasonable based on workers' own driving experience on a 7-point Likert Scale  ($1$ for pretty unreal/pretty unreasonable and $7$ for pretty real/pretty reasonable). The choice {\em Normal} means the rater maintains neutral attitude for the authenticity of a generated image and the reasonableness of the model prediction.
In RQ3, we used both MR non-violation cases (the model prediction change is deemed as rule compliance) and MR violation cases (the model prediction change is deemed rule violation) to investigate rater's general opinion towards MRs used. We didn't disclose such information (non-violation or violation cases) to raters to get unbiased rating information.

To ensure a $95\%$ confidence level with a $5\%$ confidence interval, we randomly sampled $345$ test case pairs from generated metamorphic test sets based on proposed rules except Rule 4. We did not use the test set of Rule 4 because test cases in Rule 4 are generated by the same transformation engine as in Rule 1 but with different MRs. We split $345$ test cases into 15 groups, each containing 23 test cases. We created a human intelligence task (HIT) for each group of test cases. Each task takes about 15 minutes to finish. There are 15 HITs and for each HIT, we wanted to assign 8 unique workers.  All together there shall be 120 workers. Since some workers finished multiple HITs, in the end there were 64 unique workers in total for the 15 HITs.  Since different drivers have different driving expertise and preference, we computed the inter-rater agreement using Krippendorff’s alpha. To ensure the quality of the human evaluation, we only allow workers whose task approval rate is greater than 95\% to participate. We required workers to have at least one year of driving experience. We paid each worker \$0.5 US dollars after they finished each task.

For RQ4, we compared {\tool} with prior work DeepTest and DeepRoad in two settings. In the first setting, we reproduced the common MR \textit{"If the weather changes to a rainy day, the steering angle of ego-vehicle should keep the same"} supported by DeepTest and DeepRoad. We also integrated another image-to-image translation model UGATIT~\cite{kim2019u} in {\tool} to implement the MR. Then we compared the detected violations of three methods. For DeepTest and DeepRoad, we did not find their source code or pre-trained models to generate the raining effect. Instead, we used the public code that applies OpenCV to add rain to reproduce DeepTest. For DeepRoad, we used the same network architecture as the prior work and trained the model on the A2D2 dataset. For UGATIT, we used the same dataset to train the model. In the second setting, as {\tool} supports to generate more complicated scenarios by composing multiple transformations, we evaluated whether {\tool} can detect more violations by extending test scenarios generated by DeepTest and DeepRoad. Therefore, we first applied DeepTest and DeepRoad to generate new test scenarios in rainy weather and further applied {\tool} to add a pedestrian on the roadside as more complicated scenarios.

For RQ5, we quantified the efficiency of {\tool} on generating new testing images and evaluating MR violations for different rules. For each rule, we calculated the time cost to generate the testing set and the mean cost of evaluations on three driving models. In addition, we also evaluated and compared the efficiency of {\tool} for generating test images from the same MR supported by DeepTest and DeepRoad.


\section{Evaluation results}
\label{sec:evaluation}
Section~\ref{subsection:RQ1} demonstrates how effective {\tool} is based on seven proposed rules. Section~\ref{subsection:RQ2} presents how authentic generated test cases are and how valid are detected violations in the generated test cases from the view of human raters. 
A sample of \emph{Rule 5} test images is
shown in Figure~\ref{fig:transformed_images}. To better present these images in the paper, we clipped the size of these images from $320\times 160$ to $224\times 224$.



\subsection{RQ1: The capability of {\tool} for violation detection}
\label{subsection:RQ1}

\begin{figure}
\centering
\subfloat[Original driving scene]{\includegraphics[width=.15\textwidth]{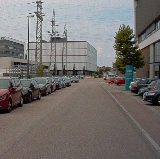}\label{fig:rule5_1}}\hfil
\subfloat[Add a pedestrian]{\includegraphics[width=.15\textwidth]{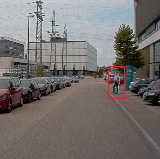}\label{fig:rule5_2}}\hfil 
\subfloat[Add a pedestrian closer]{\includegraphics[width=.15\textwidth]{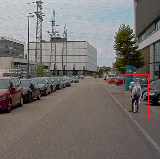}\label{fig:rule5_3}}\hfil 
\caption{Examples of source images and transformed images for Rule 4}\label{figure}
\vspace{-15pt}
\label{fig:transformed_images}
\end{figure}

\begin{table}[h]
\centering
\caption{\# violations (ratio \%) detected by seven rules}
\label{tab:exp_violation}
\scalebox{1}{
\renewcommand{\arraystretch}{1.2}
\renewcommand\tabcolsep{4pt}
\begin{tabular}{l|r|r|r}
\hline
                & \multicolumn{1}{l|}{\textbf{Epoch}} & \multicolumn{1}{l|}{\textbf{VGG16}} & \multicolumn{1}{l}{\textbf{Resnet101}} \\
                \hline
\textbf{Rule 1} & 294 (55.26\%)                       & 46 (8.65\%)                       & 68 (12.78\%)                                               \\ \hline
\textbf{Rule 2} & 244 (57.41\%)                       & 39 (9.18\%)                       & 63 (14.82\%)                                               \\ \hline
\textbf{Rule 3} & 521 (97.93\%)                       & 509 (95.68\%)                       &510 (95.86\%)                                                \\ \hline
\textbf{Rule 4} & 239 (71.34\%)                         & 115 (34.33\%)                        & 74 (22.09\%)                                                 \\ \hline
\textbf{Rule 5} & 0 (0\%)                         & 76 (15.97\%)                        & 0 (0\%)                                                   \\ \hline
\textbf{Rule 6} & 9 (1.49\%)                        & 256 (42.38\%)                        & 0 (0\%)                                                 \\ \hline
\textbf{Rule 7} & 877 (93.10\%)                       & 273 (28.98\%)                       & 226 (23.99\%)                                               \\ \hline
\end{tabular}}
\end{table}


\subsubsection{Evaluation on proposed rules}
Table~\ref{tab:exp_violation} presents experiment results of violation detected on test cases generated by seven rules. For Rules 1-4 and 7, Epoch performs worst among all five rules, with violation ratios in the range from 55.26\% to 97.93\%. The performances of VGG16 and Resnet101 are close. The violation ratios of VGG16 are lowest on Rules 1 and 2 and 
Resnet101 achieves the lowest violation ratios on Rules 4 and 7. Violation ratios of three driving models on Rule 3 are similar and all above 95\%. Epoch driving model 
performs not well for scenarios where there are pedestrians or traffic signs, or the driving scene is in the nighttime. On the other hand, VGG16 and Resnet101 
perform better.

It is notable that the violation ratios 
increase drastically on Rule 3 compared with Rule 1. 
In Rule 3, we set a more strict condition ``the deceleration should be at least 30\%" rather than simply ``the speed should decrease". 
Such customized rules are useful because they enable domain experts to test more strict requirements for driving models. For Rule 4 that combines two scenarios, it provides further evaluation for driving models. 
In Rule 1, when a driving model passes a test case, we cannot ensure whether it predicts a lower speed value because it observed a pedestrian is at the roadside. 
More violations are detected in Rule 4, which means some hidden erroneous behaviors that cannot be disclosed by simple rules are detected. Therefore, such rules combining multiple transformations (specified in the If statements) and MRs (specified in the Then statements) could be powerful tools to comprehensively evaluate driving models.  

Rules 5 and 6 aim to evaluate driving models for steering angle predictions. Epoch and Resnet101 driving models perform well.
No violation is detected for Resnet101 on Rules 5 and 6 and only 1.49\% violations are detected for Epoch on Rule 6. 
The removal of lane lines (in Rule 5) and the change of buildings (in Rule 6) do not affect the decision-making of Epoch and Resnet101. However, for VGG16, the violation ratios on Rules 5 and 6 are 15.97\% and 42.38\% respectively, which means VGG16 leverages features of lane lines and buildings 
to make predictions and is more sensitive to the change of driving environment when predicting steering angles. 

When considering the results of different driving models on different rules more comprehensively, more interesting insights could be obtained. When driving models are trained, from the original test set, we can only evaluate their performances by single metric MAE or MSE. When the values of MAE or MSE are similar in three models, it is intuitive to think their performances are similar. However, when applying different 
MTs based on different rules, we can find that their performance varies in different testing scenarios and  we are able to find the optimal driving model capable of meeting specific requirements (e.g., the model should perform well in the nighttime.). In addition, by analyzing the difference, we could find out characteristics of driving models with different architectures. In our experiment, Epoch model performs worse than other two models on rules for evaluating speed prediction (expected model predictions shall change). However, Epoch model performs better on rules for evaluating steering angle prediction  (expected model predictions shall keep the same).
This result means Epoch tends to make predictions conservatively, which may be caused by its simple network architecture. For Resnet101, it performs well on both rules for speed and steering angle predictions. It implies that this model is complicated enough to handle both tasks by learning important features for speed and steering angle predictions.




\subsubsection{Evaluation on rules with different thresholds}

Table~\ref{exp_th} shows the experiment results of applying different thresholds on rules 1, 2, 5, 6, and 7. The setting of thresholds is similar as Rule 4. For Rules 1, 2, and 7, the threshold $0$ means that the driving model should slow down and the thresholds 10\% to 50\% mean that the driving model should slow down at least 10\% to 50\% respectively. For Rules 5 and 6, the threshold 0 means that the driving model should keep the same steering angle and other thresholds mean that the driving model should deviate at least 10\% to 50\% respectively. 

The support of setting different thresholds makes {\tool} effective to meet different testing requirements (e.g., different countries may have different standards for the same traffic rule) and reveal prediction patterns of driving models.
For Rules 1 and 2, when the threshold is changed from 0 to 10\%, the number of detected violations on three models drastically increases. The result means that the decreases of speed predictions on most of testing images generated by Rules 1 and 2 are less than 10\%. With the more strict setting of the threshold, such predictions are detected as violations by {\tool}. For Rule 7, the result shows the same pattern that with the increase of the threshold, the number of detected violations also increase. For Rules 5 and 6, the detected number of violations on Epoch and Resnet101 keeps close to 0. The result means models' predictions of Epoch and Resnet101 on most of testing images from Rules 5 and 6 are close to predictions on original images.  
\begin{table}
\centering

\caption{\# violations (ratio \%) detected by rules with different thresholds}
\label{exp_th}
\arrayrulecolor{black}
\scalebox{0.85}{
\begin{tabular}{l|l|llllll} 
\hline
\rowcolor[rgb]{0.753,0.753,0.753} {\cellcolor[rgb]{0.753,0.753,0.753}}                                   & \multicolumn{1}{c|}{{\cellcolor[rgb]{0.753,0.753,0.753}}}                                  & \multicolumn{6}{c}{\textbf{Threshold }}                                                     \\ 
\hhline{>{\arrayrulecolor[rgb]{0.753,0.753,0.753}}-->{\arrayrulecolor{black}}------}
\rowcolor[rgb]{0.753,0.753,0.753} \multirow{-2}{*}{{\cellcolor[rgb]{0.753,0.753,0.753}}\textbf{Rule }}   & \multicolumn{1}{c|}{\multirow{-2}{*}{{\cellcolor[rgb]{0.753,0.753,0.753}}\textbf{Model }}} & \textbf{0} & \textbf{10\%} & \textbf{20\%} & \textbf{30\%} & \textbf{40\%} & \textbf{50\%}  \\ 
\hline
\rowcolor[rgb]{0.898,0.898,0.898} {\cellcolor[rgb]{0.898,0.898,0.898}}                                   & \textbf{Epoch}                                                                             & 55.26\%    & 96.80\%       & 97.74\%       & 97.93\%       & 98.50\%       & 98.68\%        \\
\rowcolor[rgb]{0.898,0.898,0.898} {\cellcolor[rgb]{0.898,0.898,0.898}}                                   & \textbf{VGG16}                                                                             & 86.47\%    & 88.72\%       & 93.42\%       & 95.68\%       & 96.80\%       & 97.18\%        \\
\rowcolor[rgb]{0.898,0.898,0.898} \multirow{-3}{*}{{\cellcolor[rgb]{0.898,0.898,0.898}}\textbf{Rule 1 }} & \textbf{Resnet101}                                                                         & 12.78\%    & 88.91\%       & 94.92\%       & 95.86\%       & 96.62\%       & 96.80\%        \\ 
\hline
\rowcolor[rgb]{0.753,0.753,0.753} {\cellcolor[rgb]{0.753,0.753,0.753}}                                   & \textbf{Epoch}                                                                             & 57.41\%    & 97.65\%       & 99.06\%       & 99.29\%       & 99.53\%       & 99.76\%        \\
\rowcolor[rgb]{0.753,0.753,0.753} {\cellcolor[rgb]{0.753,0.753,0.753}}                                   & \textbf{VGG16}                                                                             & 9.18\%     & 89.18\%       & 94.12\%       & 96.47\%       & 98.11\%       & 98.59\%        \\
\rowcolor[rgb]{0.753,0.753,0.753} \multirow{-3}{*}{{\cellcolor[rgb]{0.753,0.753,0.753}}\textbf{Rule 2 }} & \textbf{Resnet101}                                                                         & 14.82\%    & 94.82\%       & 96.47\%       & 96.94\%       & 97.18\%       & 97.65\%        \\ 
\hline
\rowcolor[rgb]{0.898,0.898,0.898} {\cellcolor[rgb]{0.898,0.898,0.898}}                                   & \textbf{Epoch}                                                                             & 0          & 0             & 0             & 0             & 0             & 0              \\
\rowcolor[rgb]{0.898,0.898,0.898} {\cellcolor[rgb]{0.898,0.898,0.898}}                                   & \textbf{VGG16}                                                                             & 51.89\%    & 15.97\%       & 4.83\%        & 1.68\%        & 0.42\%        & 0.21\%         \\
\rowcolor[rgb]{0.898,0.898,0.898} \multirow{-3}{*}{{\cellcolor[rgb]{0.898,0.898,0.898}}\textbf{Rule 5 }} & \textbf{Resnet101}                                                                         & 0.21\%     & 0             & 0             & 0             & 0             & 0              \\ 
\hline
\rowcolor[rgb]{0.753,0.753,0.753} {\cellcolor[rgb]{0.753,0.753,0.753}}                                   & \textbf{Epoch}                                                                             & 4.30\%     & 1.49\%        & 0.50\%        & 0             & 0             & 0              \\
\rowcolor[rgb]{0.753,0.753,0.753} {\cellcolor[rgb]{0.753,0.753,0.753}}                                   & \textbf{VGG16}                                                                             & 67.88\%    & 42.38\%       & 25.83\%       & 14.90\%       & 9.11\%        & 5.46\%         \\
\rowcolor[rgb]{0.753,0.753,0.753} \multirow{-3}{*}{{\cellcolor[rgb]{0.753,0.753,0.753}}\textbf{Rule 6 }} & \textbf{Resnet101}                                                                         & 2.81\%     & 0             & 0             & 0             & 0             & 0              \\ 
\hline
\rowcolor[rgb]{0.898,0.898,0.898} {\cellcolor[rgb]{0.898,0.898,0.898}}                                   & \textbf{Epoch}                                                                             & 93.10\%    & 99.15\%       & 99.36\%       & 99.36\%       & 99.47\%       & 99.58\%        \\
\rowcolor[rgb]{0.898,0.898,0.898} {\cellcolor[rgb]{0.898,0.898,0.898}}                                   & \textbf{VGG16}                                                                             & 28.98\%    & 68.79\%       & 86.84\%       & 92.46\%       & 94.28\%       & 95.86\%        \\
\rowcolor[rgb]{0.898,0.898,0.898} \multirow{-3}{*}{{\cellcolor[rgb]{0.898,0.898,0.898}}\textbf{Rule 7 }} & \textbf{Resnet101}                                                                         & 23.99\%    & 26.75\%       & 43.52\%       & 62.42\%       & 73.14\%       & 78.56\%        \\
\hline
\end{tabular}}
\end{table}
\subsubsection{Evaluation on complicated rules}
 
To further investigate the effectiveness of complicated rules (e.g., Rule 4) over the simple rules (e.g. Rule 1), we propose corresponding complicated rules for Rules 2, 5, 6, and 7. Specifically, for Rule 2, we generate another new image that adds a sign closer to the autonomous vehicle and check whether the speed of the driving model decrease more (called \textit{CR 1}). For Rule 5 and 6, we combine these two rules to generate testing images that remove lane lines and replace buildings with trees simultaneously and check whether the steering angle of the driving model changes (called \textit{CR 2}). For Rule 7, we combine it with Rule 1 to generate testing images that contain a pedestrian on the roadside in night driving time and check whether the speed of the driving model decrease more (called \textit{CR 3}).

Table~\ref{tab:exp1_2} shows detected violations of complicated rules on three driving models. Comparing CR 1 and Rule 2, the violation ratios increase significantly on three driving models, from 57.41\% to 78.46\% on Epoch, from 9.18\% to 56.27\% on VGG16, and from 14.82\% to 59.48\% on Resnet101 respectively. The results disclose erroneous behaviors of driving models that they cannot decelerate properly when meeting traffic signs. For CR 2, detected violations on three driving models are similar as Rules 5 and 6, which means driving models have similar behaviors for predicting steering angles when the driving environment is become more complicated. For CR 3, violations on Epoch increase about 4\% and keep similar on VGG16 and Resnet101, compared with Rule 7. Overall, complicated rules that combine multiple transformations and MRs help detect more abnormal behaviors and disclose hidden problems of driving models. 

\begin{framed}
\noindent Result 1: {\tool} can detect a large amount of abnormal behaviors of driving models using customized metamorphic relations. The customized setting of threshold and the combination of multiple test scenarios are able to cover and detect hidden problems of driving models. 
\end{framed}

\begin{table}
\centering
\caption{\# violations (ratio \%) detected by complicated rules}
\label{tab:exp1_2}
\begin{tabular}{l|l|l|l} 
\hline
              & \multicolumn{1}{c|}{\textbf{Epoch}} & \multicolumn{1}{c|}{\textbf{VGG16}} & \multicolumn{1}{c}{\textbf{Resnet101}}  \\ 
\hline
\textbf{CR 1} & 244 (78.46\%)                       & 175 (56.27\%)                       & 185 (59.48\%)                           \\ 
\hline
\textbf{CR 2} & 0 (0\%)                             & 112 (42.91\%)                       & 0 (0\%)                                 \\ 
\hline
\textbf{CR 3} & 516 (96.99\%)                       & 143 (26.88\%)                       & 132 (24.81\%)                           \\
\hline
\end{tabular}
\end{table}

\subsection{RQ2: The authenticity of transformed images generated by {\tool}}
\label{subsection:RQ2}

\begin{figure}[ht]
\centering
\includegraphics[width = .45\textwidth]{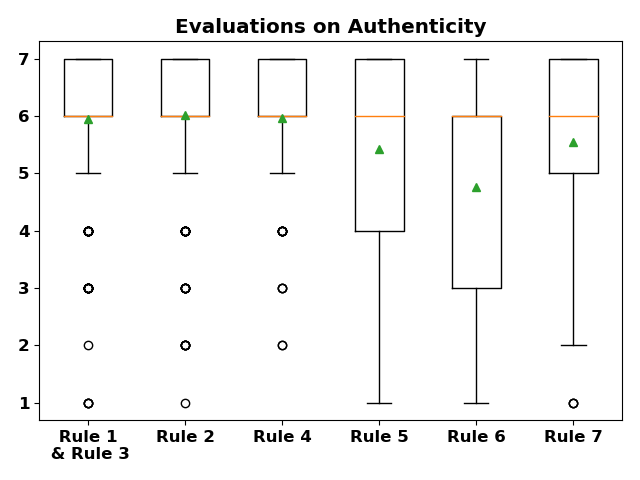}
\caption{Human assessment of the image authenticity in a 7-Point Likert Scale}
\label{fig:exp_reality}
\end{figure}

\begin{figure}[ht]
\centering
\includegraphics[width = .45\textwidth]{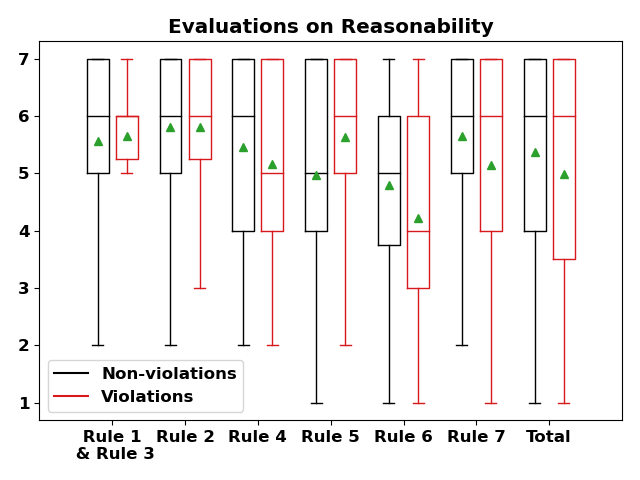}
\caption{Human assessment of the reasonability of new prediction  in a 7-Point Likert Scale.}
\label{fig:exp_reasonable}
\vspace{-5pt}
\end{figure}

64 workers were recruited in mTurk to participate in the human evaluation. Among 64 workers, 1 of them did not have driving experience; 2 had less than one year of driving experience; 8 had less than three years of driving experience, and 53 have more than three years of driving experience. We filtered out workers that did not have driving experience or had less than one year of driving experience to ensure that all results are from experienced drivers.

Figure~\ref{fig:exp_reality} demonstrates the box plot of scores on image authenticity of six transformations (Rules 1 \& 3 use the same transformation) in a 7-point Likert Scale. For each box, the green triangle is the mean, and the yellow line is the median. It can be seen for transformations of Rules 1-4, majority of ratings are in the range [6, 7], and the mean rating is about $6$. For transformations of Rules 5 and 7, the mean ratings are above 5. For the transformation of Rule 6, the mean rating is about 4.7. This relatively lower rating for Rule 6 is largely contributed by the limited capabilities of the transformation engine for this specific rule scenarios.
The results show that workers think most of the transformed images are at the level between slightly real and real, which means the quality of generated test cases is accepted. 

\begin{framed}
\noindent Result 2: Human raters consistently considered that transformed images are authentic.
\end{framed}

\subsection{RQ3: the validity of abnormal model predictions detected by {\tool}}
olds
Figure~\ref{fig:exp_reasonable} shows results on the human evaluation of the reasonability of model prediction on the transformed images. 
In the human evaluation, we randomly sampled test cases that violated MRs and did not violate MRs. 
We thus report results based on the categorization of non-violation cases and violation cases.  
For the non-violation cases, if a rater gave rating 7, it implies that the rater was in agreement with {\tool} that the model prediction on the transformed image is reasonable and model does not exhibit abnormal behavior (as the underlying MR also holds), and vice versa.  For the violation cases, if a rater gave rating 7, the detected violation is invalid because it implies that the rater was not in agreement with {\tool} that the model prediction on the transformed image exhibits abnormal behavior (as the underlying MR is violated), and vice versa.

For the non-violation cases, except for Rules 5 and 6 (with median value 5), in all other rules (with median value 6), raters in general agreed with {\tool} that when the underlying MR holds, the model does not exhibit abnormal behavior. For Rules 5 and 6, the average ratings are 4.97 and 4.80 respectively. Since both rules are steering angle basoldsed, raters might be quite conservative in determining the reasonability of model's prediction.

For the violation cases, we have an interesting observation. 
Except for Rules 6 and 7 (with average value 4 and 5 respectively and the third quartile of ratings in both rules are much lower than those on non-violation cases), in all other rules (with median value 6), raters seemed in disagreement with {\tool} that when the underlying MR violates, the model exhibits abnormal behavior. But after taking a close look at how we randomly sampled the violation and non-violation cases, we observed that for Rules 6 and 7, the violation cases have the ratios of 31.25\% and 35.29\% respectively in the total samples for each rule. And for other rules, the violation cases have much lower ratios in the rule samples for the human rater's evaluation (Rules 1 and 3 have only 5.36\% and Rule 5 has 6.90\%). 

Along with the last group of box plots for total samples show that the mean rating of reasonability on 
non-violation cases is less than that on violation cases. All these results mean that in general raters believe
violations detected by {\tool} are likely true erroneous behaviors in the underlying autonomous driving models.

To verify how consistent the ratings of different workers, we used Krippendorff's alpha to measure the agreement of rating amount multiple raters. Krippendorff's alpha is a metric generalized from other inter-rater measurements such as Fleiss' kappa~\cite{sim2005kappa}, and it is more suitable to the small size of samples and ordinal rating than Fleiss' kappa. A score less than 0 means disagreement,  and equal to 1 means perfect agreement. Therefore, we considered that the Krippendorff's alpha score should be greater than 0. We thus calculated Krippendorff's alpha score for each HIT. The agreements of authenticity (on transformation) and reasonability (on MRs) are highest at HIT 14, 0.75, and 0.60, respectively. The average scores are 0.48 and 0.37, which means the ratings of workers are fairly consistent. The result then can support the findings that test cases generated by {\tool} are authentic and detected violations are more likely actual erroneous behaviors of driving models.

\begin{framed}
\noindent Result 3: Human raters consistently considered that MR violations as potential erroneous behaviors.
\end{framed}

\subsection{RQ4: Comparison with prior work}
\label{sec:comparison}

\begin{table}[]
\centering
\caption{Comparison with DeepTest and DeepRoad for detected \# violations (ratio \%) under setting I}
\label{tab:exp_comp_1}
\begin{tabular}{l|l|l|l}
\hline
                   & \textbf{DeepTest} & \textbf{DeepRoad} & \textbf{RMT}           \\ \hline
\textbf{Epoch}     & 2 (0.21\%)        & 11 (1.17\%)       & \textbf{20 (2.12\%)}   \\ \hline
\textbf{VGG16}     & 501 (53.18\%)     & 541 (57.43\%)     & \textbf{546 (57.96\%)} \\ \hline
\textbf{Resnet101} & 0 (0\%)           & 1 (0.11\%)        & \textbf{2 (0.21\%)}    \\ \hline
\end{tabular}
\end{table}

Table~\ref{tab:exp_comp_1} shows the experiment result under the first setting above. For the same MR, {\tool} can detect more abnormal behaviors of driving models using test images generated by the more advanced image generator. Table~\ref{tab:exp_comp} shows the experiment result under the second setting. With the more complicated testing scenarios generated by {\tool}, more abnormal behaviors of three driving models are detected. Compared with DeepTest, the improvement on VGG16 achieves about $40\%$, from $18.61\%$ to $56.58\%$. Compared with DeepRoad, the biggest improvement is also on VGG16 driving model, from $15.03\%$ to $60.34\%$. The experiment result means that {\tool} can detect more abnormal behaviors of driving models by more complicated testing scenarios extended beyond prior work.


\begin{framed}
\noindent Result 4: {\tool} can detect more abnormal behaviors of driving models than prior work using more complicated test scenarios constructed by more advanced image generation techniques.
\end{framed}

\begin{table*}
\centering
\caption{Comparison with DeepTest and DeepRoad for detected \# violations (ratio \%) under setting II}
\label{tab:exp_comp}
\begin{tabular}{l|cc|cc} 
\hline
                   & \begin{tabular}[c]{@{}c@{}}\textbf{Rain added}\\\textbf{(DeepTest)}\end{tabular} & \begin{tabular}[c]{@{}c@{}}\textbf{Rain (DeepTest) and }\\\textbf{person added}\end{tabular} & \begin{tabular}[c]{@{}c@{}}\textbf{Rain added~}\\\textbf{(DeepRoad)}\end{tabular} & \begin{tabular}[c]{@{}c@{}}\textbf{\textbf{Rain (DeepRoad) and~}}\\\textbf{\textbf{person added}}\end{tabular}  \\ 
\hline
\textbf{Epoch}     & 325 (61.09\%)                                                                    & 429 (80.64\%)                                                                                & 355 (66.73\%)                                                                     & 452 (84.96\%)                                                                                                   \\ 
\hline
\textbf{VGG16}     & 99 (18.61\%)                                                                     & 301 (56.58\%)                                                                                & 80 (15.03\%)                                                                      & 321 (60.34\%)                                                                                                   \\ 
\hline
\textbf{Resnet101} & 199 (37.41\%)                                                                    & 337 (63.35\%)                                                                                & 284 (53.38\%)                                                                     & 366 (68.80\%)                                                                                                   \\
\hline
\end{tabular}
\end{table*}

\subsection{RQ5: Efficiency of {\tool}}
\begin{table}
\centering
\caption{Efficiency of {\tool} for rules}
\label{tab:exp_6}
\begin{tabular}{l|c|c|c} 
\hline
                      & \begin{tabular}[c]{@{}c@{}}\textbf{Image generation}\\\textbf{cost (s)}\end{tabular} & \begin{tabular}[c]{@{}c@{}}\textbf{Evaluation }\\\textbf{cost (s)}\end{tabular} & \textbf{Total cost (s)}  \\ 
\hline
\textbf{Rules 1 \& 3} & 137.50                                                                               & 5.58                                                                            & 143.08                   \\ 
\hline
\textbf{Rule 2}       & 131.30                                                                               & 4.80                                                                            & 137.10                   \\ 
\hline
\textbf{Rule 4}       & 279.17                                                                               & 4.80                                                                            & 283.97                   \\ 
\hline
\textbf{Rule 5}       & 196.67                                                                               & 4.78                                                                            & 201.45                   \\ 
\hline
\textbf{Rule 6}       & 195.07                                                                               & 6.08                                                                            & 201.15                   \\ 
\hline
\textbf{Rule 7}       & 126.74                                                                               & 4.37                                                                            & 131.11                   \\
\hline
\end{tabular}
\end{table}
 
Table~\ref{tab:exp_6} shows the efficiency of {\tool} on generating new testing images and evaluating MR violations of model predictions. For Rules 1-3 and Rule 7 that apply image manipulation and image-to-image translation, the cost of {\tool} on generating and filtering new test images on the testing set is about 130 seconds. For Rule 4 that needs to generate two images, the generation cost is about 280 seconds. For Rules 5 and 6 that apply a large-scale GAN model to generate new images, the image generation cost is about 195 seconds. The bottleneck is that the image generation and filtering process are applied individually on images,  because the position to add objects or pixels to remove and replace are different for each image. Overall, {\tool} can finish the test image generation and model evaluation in less than 300 seconds in the worst case of the proposed rules. When the rule is simple and low-cost image generation method is used, the time cost of {\tool} is less than 150 seconds. 

For generating test images from the same MR supported by DeepTest, DeepRoad and {\tool}, the time costs are 70.41 seconds, 118.35 seconds, and 130.90 seconds, respectively. DeepTest requires the shortest time because it does not use any image generation model to generate images. {\tool} requires about 10\% more time than DeepRoad because it uses a more advanced image generation model with higher resource consumption. In summary, RMT has marginally longer time in image generation, but it is still in the same order of magnitude as that of DeepTest and DeepRoad. Considering RMT's higher effectiveness in detecting abnormal behaviors, such marginally longer time is negligible.

\begin{framed}
\noindent Result 5: {\tool} is efficient to generate test images and evaluate driving models.
\end{framed}

\section{Threat to Validity}

\subsection{Internal Validity}
In this work, the main threats to internal validity are the completeness of proposed methods to describe driving scenes, the image generation methods and the experiment dataset. We propose to use IFTTT paradigm to describe test rules. Though the syntax of IFTTT is simple, it is suitable to cover driving scenarios in traffic rules. We propose a traffic scene ontology (Table~\ref{tab:ontology}), which is extended from a prior work evaluated on Germany highway driving scenes. The ontology is comprehensive to describe driving scenes derived from traffic rules. We define transformation and expected change inference rules (Tables~\ref{tab:predicate} and~\ref{tab:mr}) to identify image transformations and MRs. Though the transformation inference rules may not cover all possible conditions, they are powerful enough to support most of change of traffic scenes based on driving images and the relations between driving model predictions. The expected change inference rules covered all possible MRs for steering and speed predictions.  

In this work, we aim to generate high-quality driving images and apply different image generation techniques. The authenticity of images is important for driving models to make reasonable predictions. To guarantee the meaningfulness of generated images, we implement an image filtering to remove unreasonable images (e.g., a pedestrian stands on a wall). We also apply GAN-based image generation, which has been used in prior works including DeepRoad~\cite{zhang2018deeproad} to generate authentic driving images. A problem of GAN is that it is difficult to train and thus generates low quality images. To mitigate this problem, we adopt the state-of-the-art GAN architectures and augment training datasets. We use A2D2 dataset to evaluate the proposed method. Though the size of the test set (942 images) is not large, it contains traffic scenes under various conditions, including high way, rural area, and urban area. All proposed testing rules can be evaluated on the dataset. Therefore, the dataset is sufficient to evaluate abnormal behaviors that may imply faults in driving models.

\subsection{External Validity}
The threats to external validity are regarding to generalization of the proposed framework {\tool} on testing autonomous driving systems. This work focuses on the testing of image-based driving models. We select a top-performance driving model Epoch from Udacity Challenge contest and adapt two well-known and complicated classification networks VGG16 and ResNet101 to regression driving models. Thanks to the extensiveness of {\tool}, it is straightforward to add more test generation techniques, including those based on driving videos or simulations. It is thus possible to generate more complicated driving scenarios for the testing of large-scale driving systems such as Apollo~\cite{apollo}, which is our future work.      

\subsection{Construct Validity}
The threats to construct validity are related to the evaluation of driving models and the comparison with prior works. In the present work, an abnormal behavior of a driving model is considered to be detected when an MR is violated. However, in the context of autonomous driving, the abnormal behavior only implies the potential fault in a driving model. We applied qualitative user study to mitigate this issue. If the driving model's behavior is not agreed by experienced human drivers, the driving model possibly contains faults and prone to take dangerous actions.

{\tool} supports a diverse set of MRs based on traffic rules and is not restrictive for equality-based MRs, while existing techniques such as {\em DeepTest} and {\em DeepRoad} only support equality relations. Hence, in  Section~\ref{sec:comparison}, we are not arguing that this comparison is a head-to-head, fair comparison because the declarative language in {\tool} can express a super-set of what can be expressed by {\em DeepTest} and {\em DeepRoad}. We are simply assessing how much additional benefit can be provided by this improvement in MR expressiveness and extensibility. {\tool} also implements the affine transformations in {\em DeepTest} and the GAN-based transformations in {\em DeepRoad}. Therefore, {\tool} should be considered as a more general framework compared to prior work.



\section{Conclusion}
\label{sec:conclusion}
\vspace{-2pt}

In this paper, we proposed {\tool} to test the robustness of autonomous driving models. {\tool} allows users to define testing rules in natural language leveraging domain knowledge. {\tool} then implements an NLP-based rule parser for understanding testing rules, generates MRs, creates new testing images, and follow-up test sets to detect erroneous behaviors of autonomous driving models. The declarative way of defining testing rules makes {\tool} more flexible to generate different MRs and support more driving scenarios for testing.  The complete testing process is  semi-automated, with the only manual work being the user's provision of testing rules. We built a prototype of {\tool} and evaluated it with seven rules. The evaluation results showed that {\tool} effectively detected a large number of potential erroneous behaviors in three DNN-based autonomous driving models. Our qualitative study on Amazon Mechanical Turk involved experienced drivers confirmed the authenticity of generated driving scenarios and the truthfulness of detected erroneous behaviors of driving models. In addition, by comparing results of the driving model on different rule-based MT, we further evaluated performances of the model under different scenarios, which cannot be done by common testing metrics for deep learning model such as MSE. We also analyzed learned patterns of the driving model to interpret how it makes predictions.

The first prototype version of {\tool} is only used to test DNN-based driving models for predicting speed and steering angle on static road images. Currently, {\tool} does not support expressing MRs in complex driving scenarios such as vehicle overtaking, lane merging, and parallel parking. Furthermore, industry-scale autonomous driving systems, such as Baidu Apollo, consist of multiple DNN-based modules that take multi-modal input data collected from multiple kinds of sensors such as camera, Radar, and Lidar. In future work, we will extend {\tool} to test more complex, multi-model driving systems with multi-modality sensor data. In our random sampling process for RQ3, we did not implement sample balance between violation and non-violation cases leading to much less violation cases in our study. In future work, we will also implement sample balance to explore further the correlation between MRs violations and models faults.

In the future, we will extend {\tool} to support more dynamic driving scenarios and multiple modality sensor data to evaluate more complicated autonomous driving systems such as Baidu Apollo and Autoware. We will also explore the possibility to use generated test cases for improving the training of driving models.

\section*{Acknowledgment}
The authors would like to thank the anonymous reviewers for the valuable comments to improve this work. This work is in part supported by an Australian Research Council Project (DP210102447), and an ARC Linkage Project (LP190100676).


\bibliographystyle{IEEEtran}
\bibliography{IEEEabrv,references.bib}

\begin{IEEEbiography}[{\includegraphics[width=1in,height=1.25in,clip,keepaspectratio]{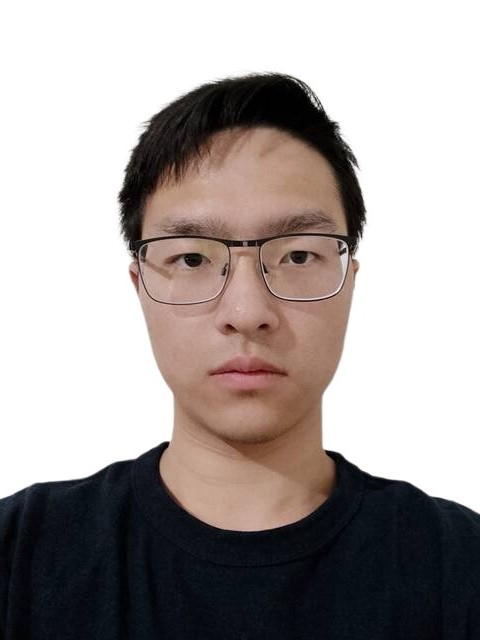}}]{Yao Deng} received the Bachelor degree of Information Technology from Deakin University, Australia in March 2018, the Bachelor degree of Software Engineering from Southwest University, China in July 2018, and the Master of Research degree in Computing from Macquarie University, Australia in 2020. He is currently pursuing his PhD degree at Macquarie University. His current research interests include machine learning, adversarial attacks and defenses, and testing on autonomous driving systems.
\end{IEEEbiography}

\begin{IEEEbiography}[{\includegraphics[width=1in,height=1.25in,clip,keepaspectratio]{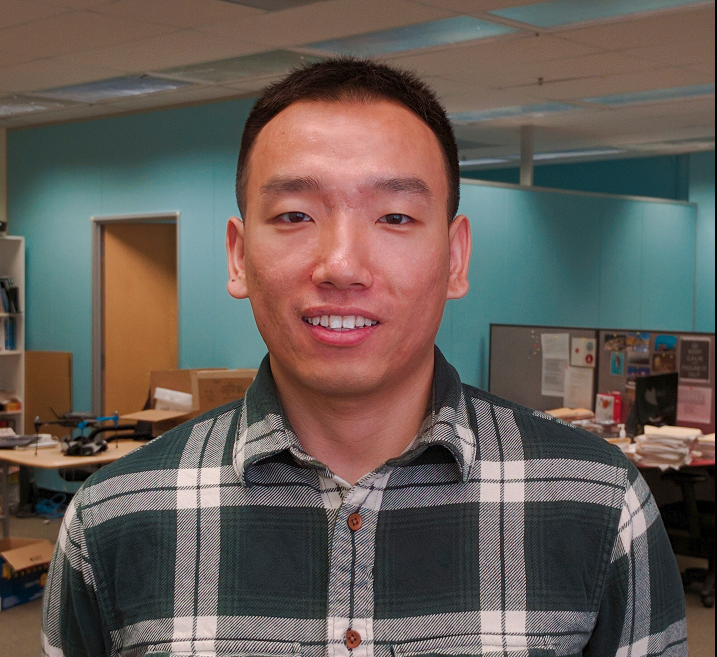}}]{Tianyi Zhang} is a Tenure-Track Assistant Professor in Computer Science at Purdue University. Prior to that, he was a Postdoctoral Fellow at Harvard University. He worked to build interactive systems that help domain experts explore and make sense of large collections of complex data, e.g., health records and code corpora. He received his Ph.D. from UCLA in 2019. His research is in the intersection of Human-Computer Interaction, Software Engineering, and AI.
\end{IEEEbiography}

\begin{IEEEbiography}[{\includegraphics[width=1in,height=1.25in,clip,keepaspectratio]{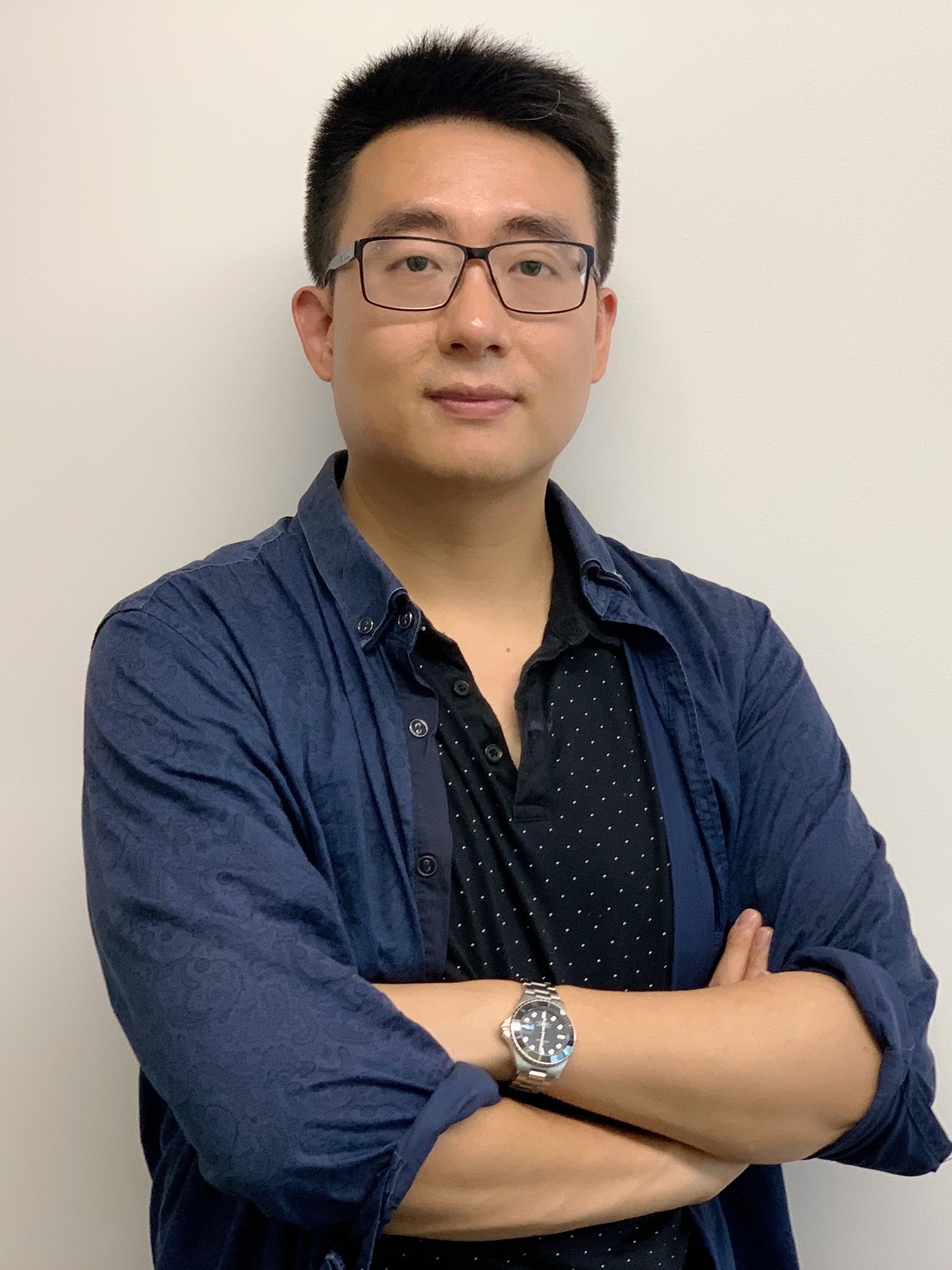}}]{Xi Zheng} received the Ph.D. in Software Engineering from UT Austin in 2015. From 2005 to 2012, he was the Chief Solution Architect for Menulog Australia. He is currently the Director of Intelligent Systems Research Group (ITSEG.ORG), Senior Lecturer (aka Associate Professor US) and Deputy Program Leader in Software Engineering, Macquarie University, Australia.  His research interests include Internet of Things, Intelligent Software Engineering, Machine Learning Security, Human-in-the-loop AI, and Edge Intelligence. He has secured more than $1$ million competitive funding in Australian Research Council (Linkage and Discovery) and Data61 (CRP) projects on safety analysis, model testing and verification, and trustworthy AI on autonomous vehicles.  He also won a few awards including Deakin Industry Researcher (2016) and MQ Earlier Career Researcher (Runner-up 2020).  He has a number of highly cited papers and best conference papers. He serves as PC members for CORE A* conferences including FSE (2022) and PerCom (2017-2022). He also serves as the PC chairs of IEEE CPSCom-2021, IEEE Broadnets-2022 and associate editor for Distributed Ledger Technologies.
\end{IEEEbiography}

\begin{IEEEbiography}[{\includegraphics[width=1in,height=1.25in,clip,keepaspectratio]{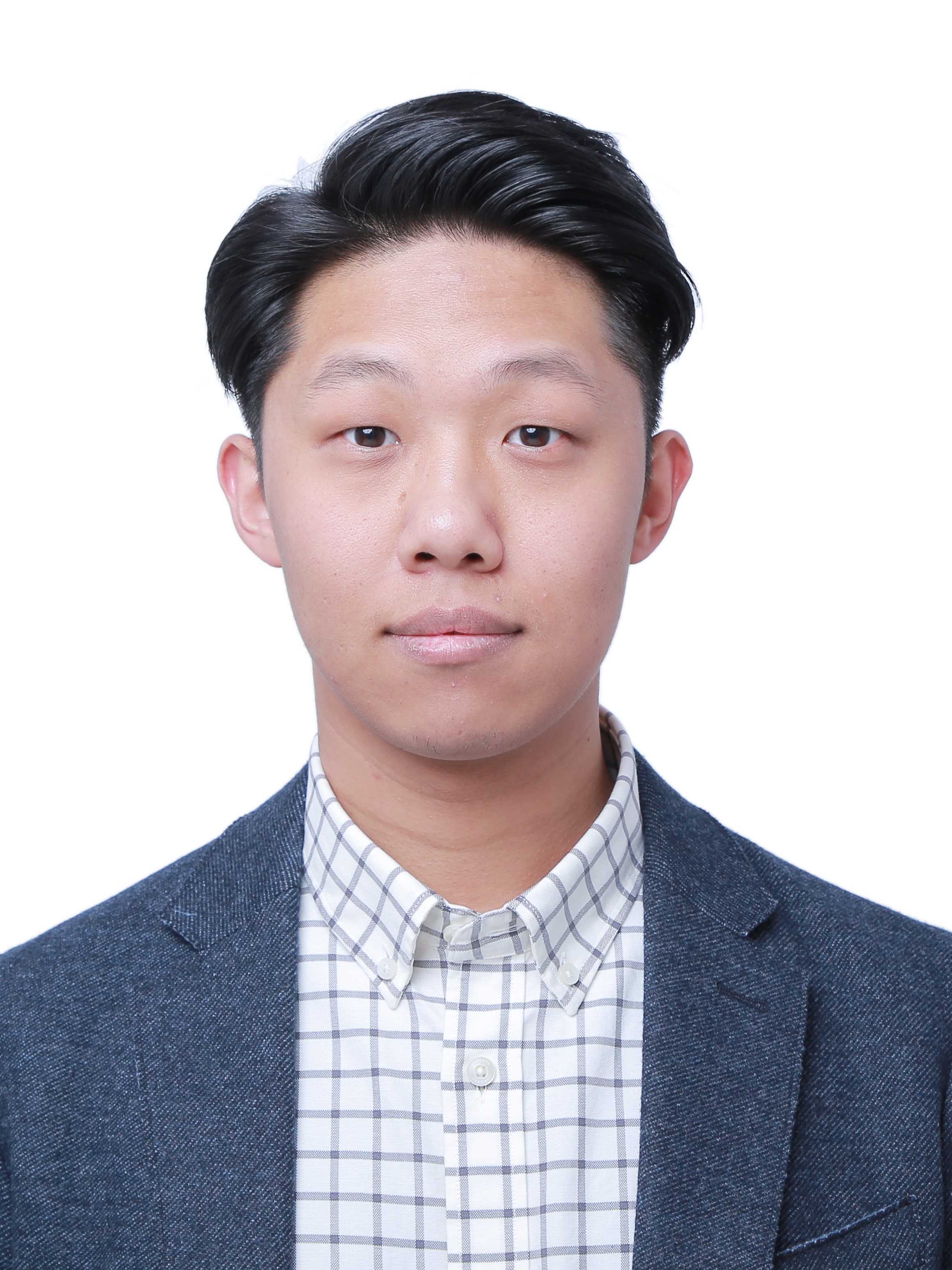}}]{Guannan Lou}
  
 received the Bachelor degree of Information Technology from Deakin University, Australia in March 2018, the Bachelor degree of Software Engineering from Southwest University, China in July 2018, and the Master of Data Science from Sydney University, Australia in 2020. His research interests include machine learning, machine learning security, metamorphic testing and natural language processing.
\end{IEEEbiography}

\begin{IEEEbiography}[{\includegraphics[width=1in,height=1.25in,clip,keepaspectratio]{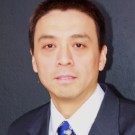}}]{Huai Liu} received the BEng degree in physioelectronic technology, the MEng degree in communications and information systems, both from Nankai University, China, and the PhD degree in software engineering from the Swinburne University of Technology, Australia. He is a senior lecturer in the Department of Computing Technologies, Swinburne University of Technology, Melbourne, Australia. He has worked as a lecturer at Victoria University and a research fellow at RMIT University. His current research interests include software testing, cloud computing, and end-user software engineering. He is a senior member of the IEEE.
\end{IEEEbiography}

\begin{IEEEbiography}[{\includegraphics[width=1in,height=1.25in,clip,keepaspectratio]{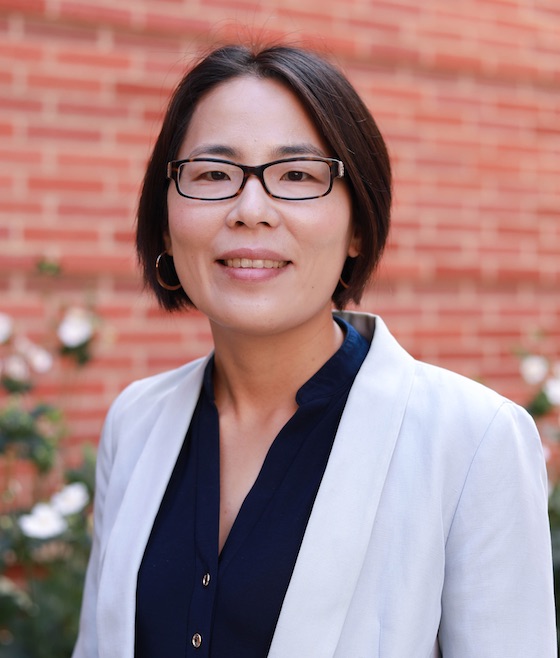}}]{Miryung Kim}
is a Professor in Computer Science at UCLA and a Director of Software Engineering and Analysis Laboratory. She helped define the new area of Software Engineering for Data Analytics (SE4DA and SE4ML). She received her BS from KAIST and MS and PhD from University of Washington under the supervision of David Notkin. She was previously an Assistant Professor at the University of Texas at Austin, moved to UCLA as an Associate Professor with tenure in 2014, and was promoted to a Full Professor in 2019. She also spent time as a visiting researcher at Microsoft Research.
\end{IEEEbiography}

\begin{IEEEbiography}[{\includegraphics[width=1in,height=1.25in,clip,keepaspectratio]{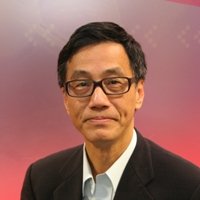}}]{Tsong Yueh Chen}
received the BSc and MPhil degrees from The University of Hong Kong, the MSc degree and DIC from the Imperial College of Science and Technology, London, U.K., and the PhD degree from The University of Melbourne, Australia. He is currently a Professor of Software Engineering in the Department of Computer Science and Software Engineering, Swinburne University of Technology, Australia. Prior to joining Swinburne, he has taught at The University of Hong Kong and The University of Melbourne. He is the inventor of metamorphic testing and adaptive random testing. His current research interests include software testing, debugging, and program repair.
\end{IEEEbiography}

\end{document}